\documentclass[11pt,a4paper]{article}
\usepackage{jheppub}

\usepackage{multirow, graphicx,amssymb,url,mathrsfs,amsmath}
\usepackage{wrapfig,boxedminipage,setspace,subfigure,epsfig}
\usepackage{amsxtra,amstext,latexsym,dsfont,amsfonts}
\usepackage{color,eucal}
\usepackage[dvipsnames]{xcolor}
\usepackage{float}
\usepackage{slashed,comment}
\usepackage{kotex}
\usepackage{tensor}
\usepackage{mathtools}
\usepackage{indentfirst}

\usepackage{array} 
\usepackage[section]{placeins}

\title{Pole-skipping for massive fields and the Stueckelberg formalism}

\author[a]{Wen-Bin Pan,}
\author[b,c]{Ya-Wen Sun,}
\author[b]{and Yuan-Tai Wang}

\emailAdd{panwb@ihep.ac.cn}
\emailAdd{yawen.sun@ucas.ac.cn}
\emailAdd{wangyuantai19@mails.ucas.ac.cn}

\affiliation[a]{Institute of High Energy Physics, Chinese Academy of Sciences,\\19B Yuquan Road, Shijingshan District, Beijing 100049, China}
\affiliation[b]{School of Physical Sciences, University of Chinese Academy of Sciences,\\ Zhongguancun east road 80, Beijing 100190, China}
\affiliation[c]{Kavli Institute for Theoretical Sciences, University of Chinese Academy of Sciences,\\ Zhongguancun east road 80, Beijing 100190, China}

\abstract{Pole-skipping refers to the special phenomenon that the pole and the zero of a retarded two-point Green's function coincide at certain points in momentum space.
We study the pole-skipping phenomenon in holographic Green's functions of boundary operators that are dual to massive $p$-form fields and the dRGT massive gravitational fields in the AdS black hole background. Pole-skipping points for these systems are computed using the near horizon method. 
The relation between the pole-skipping points of massive fields and their massless counterparts is revealed.
In particular, as the field mass $m$ is varied from zero to non-zero, the pole-skipping phenomenon undergoes an abrupt change with doubled pole-skipping points found in the massive case.  
This arises from the breaking of gauge invariance due to the mass term and the consequent appearance of more degrees of freedom.
We recover the gauge invariance using the Stueckelberg formalism by introducing auxiliary dynamical fields.
The extra pole-skipping points are identified to be associated with the Stueckelberg fields.
We also observe that, as the mass varies, some pole-skipping points of the wave number $q$ may move from a non-physical region with complex $q$ to a physical region with real $q$. 
}


\arxivnumber{}

\begin{document}
 \maketitle
\flushbottom

	
\section{Introduction}\label{sec1}

The holographic correspondence has established a profound connection between quantum gravity in asymptotically anti-de Sitter (AdS) space and a quantum field theory (QFT) living on the AdS boundary \cite{Maldacena:1997re}. 
In this framework, as the boundary theory approaches a specific strong coupling limit, the bulk theory simplifies to classical gravity coupled with matter fields. 
Within this limiting regime, the data of QFT is encoded within the classical geometry and matter fields. 
This duality serves as a tool, enabling us to compute the characteristics of strongly coupled quantum field theories, such as the retarded Green's functions which represent the near equilibrium behavior of the system \cite{Gubser:1998bc,Witten:1998qj,Kovtun:2005ev}.

In recent years, a universal property of holographic Green's functions known as ``pole-skipping'' has been investigated intensely.
This term refers to the feature that the pole and the zero of a Green's function may intersect in complex momentum space at some specific points $(\omega_*,q_*)$ where the Green's function becomes undefined, hence denominated as ``pole-skipping''.
The pole-skipping phenomenon was first discovered in the study of chaotic properties of quantum many-body systems, with the leading order skipped pole of energy density two-point function, which lies at the upper half complex plane of $\omega$, related to the characteristic parameters (the Lyapunov exponent and the butterfly velocity) of the out-of-time ordered correlation function (OTOC) \cite{Grozdanov:2017ajz, Blake:2017ris, Blake:2018leo} \footnote{Earlier studies \cite{Amado:2007yr,Amado:2008ji} also found a similar property for the Green's function, termed as "residues of correlators".}.

Apart from the chaotic pole-skipping point, there are countably infinite pole-skipping points located at $\omega=0$ or the lower half complex plane of $\omega$, which do not seem to be related to quantum chaos.
To obtain all the pole-skipping points, a holographic near horizon method was developed in \cite{Blake:2019otz}.
It is found that the pole-skipping information of Green's function of boundary QFT can be extracted simply from the properties of near-horizon probe fields.
Specifically, if the frequency and wave number take some special value $(\omega_*,q_*)$, the field would have a family of infinitely many ingoing solutions near the horizon of the AdS black hole.
In that special case, the retarded Green's function of the boundary QFT cannot be defined at $(\omega_*,q_*)$ according to the holographic dictionary \cite{Zaanen:2015oix}.
Using this method, pole-skipping points of scalar, vector, tensor, and spinor fields in various holographic space time background have been investigated \cite{Blake:2019otz, Natsuume:2019sfp,Natsuume:2019xcy,Ahn:2019rnq,Natsuume:2019vcv,Wu:2019esr,Ceplak:2019ymw,Natsuume:2020snz,Ahn:2020bks,Ahn:2020baf,Ceplak:2021efc,Blake:2021hjj,Wang:2022xoc,Jeong:2023rck} \footnote{We note that the near horizon method of pole-skipping can be extended to non-AdS \cite{Grozdanov:2023txs} and even non-black hole space times \cite{Natsuume:2023hsz,Natsuume:2023lzy}. It also infers non-uniqueness of scattering amplitudes at special points \cite{Natsuume:2021fhn}}.
The pole-skipping values of frequency $\omega_*$ generically lie at $\omega_n = -2\pi i(n-l) T, (n=1,2,3...)$, where $l$ is the spin of the field and $T$ is the temperature of background AdS black hole \cite{Wang:2022mcq,Ning:2023ggs}, while $q_*$ depend on the space time as well as the type of the field. 
The physical interpretation or implication for these non-chaotic pole-skipping points include the ``exchange effects" of corresponding fields \cite{Kim:2020url}, a universal bound on transport coefficients (such as diffusion constant) in many-body systems \cite{Jeong:2021zhz}, an order parameter to probe a quantum critical point \cite{Abbasi:2023myj}, and the constraints and reconstruction of quasi-normal modes \cite{Abbasi:2020xli,Grozdanov:2020koi,Grozdanov:2023tag}
(see \cite{Grozdanov:2018kkt,Li:2019bgc,Abbasi:2019rhy,Das:2019tga,Abbasi:2020ykq,Ramirez:2020qer,Choi:2020tdj,Sil:2020jhr,Yuan:2020fvv,Kim:2021xdz,Yuan:2021ets,Jeong:2022luo,Amano:2022mlu,Loganayagam:2022teq,Baishya:2023nsb,Yuan:2023tft,Amrahi:2023xso,Jeong:2023ynk,Baishya:2023xbj,Baishya:2023ojl} for more developments and applications of pole-skipping.).

In most of the previous studies, the pole-skipping points are computed by the near horizon analysis of the equations of motion for gauge invariant variables (see \cite{Ahn:2024aiw} for a recent comprehensive work that employs gauge-invariant variables.).
This is convenient when a gauge symmetry is present for the bulk field because every gauge invariant variable is associated with a dynamical degree of freedom and obeys a single equation of motion in simple cases such as (the longitudinal and transverse channel of) massless vector fields and (the sound, diffusive and tensor channel of) metric field fluctuations. 
However, for massive gauge fields, the gauge symmetry is broken and more degrees of freedom emerge. 
In other words, the number of degrees of freedom does not change continuously as the mass $m$ varies from zero to non-zero.
Therefore the pole-skipping analysis is more complicated in such cases. 
Gauge invariance can be recovered by the Stueckelberg formalism \cite{Ruegg:2003ps} by introducing auxiliary dynamical fields, but more gauge invariant variables are involved and one has to deal with a set of coupled equations.


In this work, we study the effect of ``mass-induced" gauge symmetry breaking on the pole-skipping phenomenon.
In particular, we investigate the pole-skipping points associated with the following two kinds of massive gauge fields.
The first one is the massive $p$-form field which is the generalization of a massive electromagnetic field ($1$-form).
The pole-skipping of $p$-form fields has been studied in a previous work \cite{Wang:2022xoc}, but the focus was mainly on the massless case. 
We give a more comprehensive treatment to the massive case in this paper.
The other type of massive field is the massive gravitational field in a ghost-free massive gravity, proposed by de Rham, Gabadadze, and Tolley (dRGT) \cite{deRham:2010ik,deRham:2010kj}, which was utilized to describe momentum dissipation in AdS/CFT \cite{Vegh:2013sk,Blake:2013bqa,Davison:2013jba}.
The work \cite{Liu:2020yaf} studied pole-skipping in topological massive gravity in three dimensions.
The massive gravity part of the present work can be seen as a higher dimensional generalization of the mass effect on the pole-skipping of that system.

Our main aim is to provide a systematic approach to analyse the pole-skipping phenomenon in holographic models that a gauge symmetry is broken by the mass. 
All possible channels of the two kinds of massive fluctuations are considered and a near-horizon method that takes advantage of matrices is used to calculate the pole-skipping from a set of coupled equations.
Once a non-zero mass term is turned on, we find that the number of skipped poles would be doubled in the longitudinal channel of $p$-form fields as well as the diffusive and sound channel of metric fluctuations.
We demonstrate further that the doubled skipped poles arise from the extra degree of freedom of the auxiliary field in the Stueckelberg formalism.
We also examine whether the skipped poles of each massive channel lie in the physical region (i.e. the wave number $q_*$ of the skipped poles is real) as the mass $m$ varies.
It turns out that this feature of skipped poles indeed depends on the mass.

The remaining content of this paper is organized as follows. 
In section \ref{sec2}, we briefly review the pole-skipping of the massless correlators with the near horizon analysis of the $1$-form field in section \ref{sec2.1}, the $p$-form field in section \ref{sec2.2} and the gravitational field in section \ref{sec2.3}.
In section \ref{sec3}, we study the role of field mass on pole-skipping properties and explore the connection of pole-skipping between massless fields and their massive counterparts.
Section \ref{sec3.1}, \ref{sec3.2} and \ref{sec3.3} are devoted to massive $1$-form fields, massive $p$-form fields and dRGT massive gravitational fields respectively.
In section \ref{sec4}, we discuss whether the skipped poles are physical and summarize this paper.
Appendix \ref{appendixA} is devoted to an alternative computing method which involve gauge invariant variables.
Finally, we also show the pole-skipping results of $p$-form fields in a AdS-Schwarzschild black hole in appendix \ref{appendixB}.

\section{Review of pole-skipping for massless fields}\label{sec2}

In this section, we review the pole-skipping of massless fields, including the $p$-form fields and metric fluctuations.
We also define the notation and terminology to pave the way for further calculations.



\subsection{Massless $1$-form fields}\label{sec2.1}

We first review some of the basics by considering the current-current retarded Green's functions and calculating their pole-skipping holographically following \cite{Blake:2019otz}.
We set the ($d+2$)-dimensional bulk metric to be 
\begin{align}\label{metric}
    d s^{2}=-r^{2} f(r) d t^{2}+\frac{1}{r^{2} f(r)} d r^{2}+h(r) d \vec{x}^{2}
\end{align}
We assume that the geometry \eqref{metric} is an asymptotically AdS black hole with $r\to \infty$ corresponding to the boundary and the hypersurface $r=r_0$ constitutes a planar horizon.
The functions $f(r)$ and $h(r)$ fulfill certain boundary conditions at the AdS boundary ($f(r\rightarrow\infty)\sim 1$ and $h(r\rightarrow\infty)\sim r^2$) and the event horizon ($f(r_0)=0$ and $h(r)$ is regular at $r_0$).
We do not restrict the geometry \eqref{metric} to be Schwarzschild-AdS, but assume it can be a more general geometry supported by additional matter (i.e. \eqref{metric} is a solution of Einstein equation coupled with matter).
The Hawking temperature of this black hole is $T=\frac{r_0^2 f'(r_0)}{4\pi}$.
The Eddington-Finkelstein (E-F) coordinates are convenient for us to consider the ingoing solutions of the fields which determine the retarded Green's function on the boundary.
The metric in the ingoing E-F coordinates becomes
\begin{align}\label{EFmetric}
    d s^{2}=-r^{2} f(r) d v^{2}+2 d v d r+h(r) d \vec{x}^{2},
\end{align}
where we have defined $v=t+r^*$, $\frac{d r_{*}}{d r}=\frac{1}{r^{2} f(r)}$ and $d \vec{x}^{2}=dx^2+dy_1^2+...+dy_{d-1}^2$.
The ingoing E-F coordinates are used throughout the paper.

To compute the current-current retarded Green's functions, we add a probe gauge field $A_\mu$ to the above background geometry
\begin{align}
    S=-\frac{1}{4}\int d^{(d+2)}x \sqrt{-g} Z(\Phi) F^{\mu \nu}F_{\mu \nu},
\end{align}
where $F_{\mu \nu}=\partial_\mu A_\nu-\partial_\nu A_\mu$ and it is invariant under the gauge transformation $A_\mu \to A_\mu+\partial_\mu \Lambda$.
Note that a dilaton field $\Phi(r)$ depending only on the radial coordinate is added to the conventional electromagnetic action, the same as in \cite{Blake:2019otz}.
It is convenient to focus on the equation of motion for gauge invariants, instead of the gauge field $A_\mu$.
The equation of motion is
\begin{align}\label{eom1F}
    \nabla_\mu (Z(r) F^{\mu \nu})=\frac{1}{\sqrt{-g}}\partial_\mu (\sqrt{-g} Z(r) g^{\mu \rho}g^{\nu \sigma} F_{\rho \sigma})=0.
\end{align}
$F$ is an exact 2-form, so it also satisfies the constraint $dF=0$ or equivalently
\begin{align}\label{constraint1F}
    \partial_{[\rho} F_{\mu \nu]}=\partial_{\rho} F_{\mu \nu}+\partial_{\mu} F_{\nu \rho}+\partial_{\nu} F_{\rho \mu}=0.
\end{align}
In the ingoing EF coordinates, we assume that the fluctuations propagate toward the $x$ direction with wave number $q$, i.e. the gauge invariants have the form of
\begin{align}
    F_{\mu \nu}(v,r,x,y_{j}) \to F_{\mu \nu}(r) e^{-i\omega v+iqx}.
\end{align}

According to the behavior of each component of $A_\mu$ under rotational transformations around the $x$ axis, we can divide them into two channels: the longitudinal channel ($A_v$, $A_r$, $A_x$) and the transverse channel ($A_j$ with $j=1,...,d-1$ corresponding to the transverse coordinate $y_j$).
Each channel corresponds to some gauge invariant variables, each of which is a combination of the gauge field components in that channel \cite{Kovtun:2005ev}.
The longitudinal channel corresponds to $F_{vx}=-i\omega A_x-i q A_v$, $F_{vr}=-i\omega A_r-\partial_r A_v$ and $F_{rx}=\partial_r A_x-i q A_r$, while the transverse channel includes $F_{vj}=-i\omega A_j$, $F_{rj}=\partial_r A_j$ and $F_{xj}=i q A_j$.
The two channels in both the equation of motion \eqref{eom1F} and the constraint \eqref{constraint1F} decouple.

In the longitudinal channel, the equations of motion for $F_{vr}$, $F_{vx}$ and $F_{rx}$ can be reduced into a single equation for $F_{vx}(r)$
\begin{equation}\label{master_long_1}
    \frac{d}{d r}\left[\frac{h^{d / 2} Z}{\omega^{2} h-q^{2} r^{2} f}\left(i r^{2} f \frac{d F_{vx}}{dr}+\omega F_{vx}\right)\right]+\frac{h^{d / 2-1} Z}{\omega^{2} h-q^{2} r^{2} f}\left( \omega h \frac{d F_{vx}}{dr}-i q^{2} F_{vx}\right)=0。
\end{equation}
Other gauge invariant variables can be easily obtained once the equation \eqref{master_long_1} is solved, so it is called the master equation for the longitudinal channel.
In the transverse channel, the equations of motion and constraints for $F_{vj}$, $F_{rj}$ and $F_{xj}$ can also be reduced to a single master equation
\begin{equation}\label{master_tran_1}
    F_{xy}''+\left(\frac{2}{r}+\frac{r^2 f'-2i\omega}{r^2 f}+\frac{(d-2)h'}{2h}+\frac{Z'}{Z}\right)F_{xy}'-\left(i\frac{(d-2)\omega h'-2i q^2}{2r^2 f h}+\frac{\omega Z'}{r^2 f Z}\right)F_{xy}=0.
\end{equation}
In each channel, the master equation can be rearranged into a second-order ordinary differential equation with the form of
\begin{equation}\label{masterF}
    F''(r)+a(r)F'(r)+b(r)F(r)=0,
\end{equation}
where $F(r)$ represents a gauge invariant variable ($F_{vx}(r)$ for the longitudinal channel and $F_{xy}(r)$ for the transverse channel).

To compute the current-current retarded Green's functions $C_{\mu \nu}(\omega,q)$ \footnote{$C_{\mu \nu}(\omega,q)$ is defined as the Fourier transformation of the current-current retarded Green's functions in the coordinate space
$C_{\mu \nu}(x-y)=-i \theta\left(x^{0}-y^{0}\right)\left\langle\left[J_{\mu}(x), J_{\nu}(y)\right]\right\rangle$.}, one first need to solve the equation \eqref{masterF} (for each channel) near the horizon $r_0$.
In general, there are two distinguished solutions characterized by whether it falls into the horizon (ingoing) or propagate out from the horizon (outgoing).
We solve for the ingoing solution since retarded Green's functions are needed.
After the ingoing solution is obtained, one can extend it to the AdS boundary and its near-boundary expansion has the form of
\begin{align}\label{bdryexp}   
    F(r)=A(\omega,q) r^{-\Delta_{-}}(1+\cdots)+B(\omega,q) r^{-\Delta_{+}}(1+\cdots)
\end{align}
$\Delta_{+}$ and $\Delta_{-}$ are the scaling dimensions of dual current operator $J^\mu$ and source $A^{(0)}_{\mu}$ (boundary value of bulk field) on the boundary QFT respectively \footnote{For the $1$-form field, we have $\Delta_{+}=d$ and $\Delta_{-}=1$.}.  
The retarded Green's functions can be computed by the GKPW relation, and after holographic renormalization \cite{deHaro:2000vlm,Skenderis:2002wp}, the result is \footnote{see \cite{Kovtun:2005ev} for more details.}
\begin{align}\label{JJ1}
    C_{tt}= \frac{ \alpha q^{2}}{\omega^{2}-q^{2}} \frac{B_l (\omega,q)}{A_l(\omega,q)}, \quad 
    C_{tx}=C_{xt}=\frac{\alpha \omega q}{\omega^{2}-q^{2}} \frac{B_l(\omega,q)}{A_l(\omega,q)}, \quad 
    C_{xx}=\frac{\alpha \omega^{2}}{\omega^{2}-q^{2}} \frac{B_l(\omega,q)}{A_l(\omega,q)}
\end{align}
and
\begin{align}\label{JJ2}  
    C_{jj}=\alpha \frac{B_t(\omega,q)}{A_t(\omega,q)},
\end{align} 
where $\alpha$ is a normalization constant, $A_l(\omega,q)$ and $B_l(\omega,q)$ denote the coefficients for the longitudinal channel while $A_t(\omega,q)$ and $B_t(\omega,q)$ denote the coefficients for the transverse channel.
We note that all the Green's functions are proportional to the ratio of $\frac{B(\omega,q)}{A(\omega,q)}$.

However, in cases where the pair $(\omega,q)$ takes some special values, the ingoing boundary condition cannot determine the solution of the equation of motion uniquely (up to a global constant) and there would be two independent solutions that both fulfill the ingoing boundary condition.
It turns out that the boundary retarded Green's functions cannot be uniquely defined at these special points of momentum space, due to the non-uniqueness of $B(\omega,q)$ and $A(\omega,q)$.
In the following, we show how to obtain special points of $(\omega,q)$ through the near-horizon solution of eq. \eqref{masterF}.

To obtain the near horizon solution, we assume the solution of master eq. \eqref{masterF} has the following expansion near the horizon $r_0$
\begin{align}\label{expphi}
	F(r)=(r-r_0)^\lambda \sum_{j=0}^{\infty} F_{j}\left(r-r_{0}\right)^{j}=(r-r_0)^\alpha (F_{0}+F_{1}\left(r-r_{0}\right)+\ldots),
\end{align}
where $\lambda$ is a constant to be determined by the equation of motion and $F_{j}$ are coefficients to be determined.
The two coefficients $a(r)$ and $b(r)$ also have expansions of the form,
\begin{equation}\label{expab}
	\begin{aligned}
		a(r)=\frac{a_0}{r-r_0}+a_1+a_2 (r-r_0)+ O(r-r_0)^2,\\
		b(r)=\frac{b_0}{r-r_0}+b_1+b_2 (r-r_0)+ O(r-r_0)^2,
	\end{aligned}
\end{equation}
with $a_i$ and $b_i$ depending on $\omega$ and $q$.
Put the two near horizon expansion \eqref{expphi}, \eqref{expab} into eq. \eqref{masterF}, and we find the series expansion of eq. \eqref{masterF} can be written as
\begin{equation}\label{expeom}
\begin{aligned}
	\mathcal{S} &= \mathcal{S}_0 (r-r_0)^{\lambda-2}+ \mathcal{S}_1 (r-r_0)^{\lambda-1} + \mathcal{S}_2 (r-r_0)^{\lambda} + \cdots = 0,
\end{aligned}
\end{equation}
where 
\begin{equation}
	\begin{aligned}
		\mathcal{S}_0&=\lambda(\lambda-1+a_0)F_{0}= 0,\\
		\mathcal{S}_1&=(\lambda+1)(\lambda+a_0)F_{1}+(\lambda a_1+b_0)F_{0}= 0,\\
		\mathcal{S}_2&=(\lambda+2)(\lambda+1+a_0)F_{2}+((\lambda+1) a_1+b_0)F_{1}+(\lambda a_2+b_1)F_{0}= 0,\\
		\cdots
	\end{aligned}
\end{equation}

Next, we solve for the coefficients $F_i,i=1,2\cdots$ order by order (i.e. represent $F_i,i=1,2\cdots$ by $F_0$).
According to the zeroth order equation,
\begin{equation}
    \mathcal{S}_0=\lambda(\lambda-1+a_0)F_{0}=0,
\end{equation}
$\lambda$ has two possible values: $\lambda_1=0$ and $\lambda_2=1-a_0$, where $a_0=1-\frac{i\omega}{2\pi T}$ for both channels of the $1$-form field.
For general value of $\omega$, the solution with leading term $(r-r_0)^{\lambda_1}$ is an ingoing solution while the other leading behavior $(r-r_0)^{\lambda_2}$ corresponds to an outgoing solution.
Substituting in the ingoing solution, the first order equation becomes
\begin{align}\label{S1}
	 \mathcal{S}_1 &=a_0 F_{1}+b_0 F_{0}=0.
\end{align}
In general, $F_1$ can be expressed by $F_0$ from \eqref{S1}, then higher order coefficients $F_j$ (j=2,3,...) can be solved iteratively.
Since $F_j$ depends only on $F_0$ for $j=1,2,3,...$, there is only one ingoing solution, which is proportional to the constant $F_0$.

However, there exists one point $(\omega_*,q_*)$ that satisfies the condition $a_0(\omega_*,q_*)=b_0(\omega_*,q_*)$\\$=0$, which is special.
In this case, $F_0$ and $F_1$ are both undetermined free parameters.
If we proceed the iteration, $F_j$ with $(j=2,3,...)$ would depend both on $F_0$ and $F_1$.
The most general solution of the equation of motion can be written as a linear combination of the ingoing (leading behavior $F_0 (r-r_0)^0$) and outgoing (leading behavior $F_1 (r-r_0)^1$) solutions
\begin{align}
	F=F_{0}\left[1+c_{1}\left(r-r_{0}\right)+\ldots\right]+F_{1}\left(r-r_{0}\right)\left[1+d_{1}\left(r-r_{0}\right)+\ldots\right].
\end{align}
Remember that $a_0(\omega_*,q_*)=0$ at the special point, which leads to $\lambda_2$ being $1$. 
It follows that this linear combination is ingoing overall so long as $F_{0}\ne0$ and no matter what the value of $F_{1}$ is. 
This is because the outgoing solution becomes a part of series expansion (starting from the order $r-r_0$) of the ingoing solution and whether a function of $r$ is ingoing or outgoing depends only on its leading behavior near the horizon.
The ingoing solution is therefore not unique.  
This amounts to giving us a non-unique retarded Green's function of boundary operators.
For instance, suppose $F ^ {(1)} $ and $F ^ {(2)} $ are two ingoing solutions for eq. \eqref{masterF}.
When extended to the AdS boundary, their asymptotic expansions near the AdS boundary are
\begin{equation}
	\begin{aligned}
		F^{(1)}=A_1(\omega,q) r^{-\Delta_{-}}(1+\cdots)+B_1(\omega,q) r^{-\Delta_{+}}(1+\cdots),\\
		F^{(2)}=A_2(\omega,q) r^{-\Delta_{-}}(1+\cdots)+B_2(\omega,q) r^{-\Delta_{+}}(1+\cdots),
	\end{aligned}
\end{equation} respectively.
According to the aforementioned GKPW relation and holographic dictionary, one would obtain two retarded Green's functions
\begin{equation}
\begin{aligned}    
	G_1^R(\omega,q)\propto \frac{B_1(\omega,q)}{A_1(\omega,q)},\\
	G_2^R(\omega,q)\propto \frac{B_2(\omega,q)}{A_2(\omega,q)}.
\end{aligned}
\end{equation}
Moreover, since the linear combination of $F^{(1)}$ and $F^{(2)}$ is also an ingoing solution, and there is a specific linear combination such that the asymptotic expansion coefficient $B$ vanishes, the retarded Green's function obtained from this linear combination is zero. 
There is also another linear combination that gives $A=0$, and the Green's function diverges.
Therefore, the retarded Green's function is multi-valued at the special point $(\omega_*,q_*)$.
In particular, it could be both the zero and the pole of the Green's function.
Hence the special point $(\omega_*,q_*)$ is a pole-skipping point for the corresponding field. 

In general, there are also more pole-skipping points which can be derived from higher order equations ($n=2,3,4,...$)\footnote{With the assumption that for any $i=1,2,\cdots,n-1$, $\tilde{a}_{i}\ne 0$ and $\tilde{b}_{i}\ne 0$, we have substituted all $F_{2}$, $F_{3}$,...,$F_{n-1}$ by $F_0$.}
\begin{equation}
	\begin{aligned}    
		\mathcal{S}_n=\tilde{a}_{n} F_{n}+\tilde{b}_{n} F_{0}=0.
	\end{aligned}
\end{equation} 
If $\tilde{a}_{n}=\tilde{b}_{n}=0$ hold, then $F_{n}$ would become an additional free parameter. 
This would also lead to the non-uniqueness of the ingoing solution, and for the same reason, pole-skipping occurs.
Hence, the special value of $\omega$ and $q$ that makes $\tilde{a}_{n}=\tilde{b}_{n}=0$ is also a pole-skipping point.

The analysis above shows that the criterion for determining a pole-skipping point is whether there is an additional free parameter in the ingoing series solution of the equation of motion of the field. 
An additional free parameter corresponds to a one-parameter family of additional ingoing solutions, which is responsible for the multi-valuedness of the retarded Green's function.
We use this criterion to compute the pole-skipping points in the follows.




Following this procedure, the computation and results for the massless $1$-form field can be found in \cite{Blake:2019otz} and \cite{Natsuume:2019sfp}. 
In appendix \ref{appendixA}, we present an alternative computation method that is different from the previous literature and is easy to generalize to massive cases.

\subsection{Massless $p$-form fields}\label{sec2.2}

As a generalization to the electromagnetic field, we have the fully antisymmetric tensor field: the $p$-form field.
The electromagnetic field can be viewed as a $1$-form field. 
The action of a minimal coupled $p$-form field $P$ is
\begin{align}
	S=-\frac{1}{2(p+1)}\int \mathrm{d}^{(d+2)}x \sqrt{-g} Z(\Phi) (\mathrm{d} P)^2,
\end{align}
where $\mathrm{d}P$ is the field strength and is a $(p+1)$-form field which can be written explicitly as
\begin{equation}\label{dPtoP}
	\begin{aligned}
		(\mathrm{d} P)_{\alpha_{1} \cdots \alpha_{p+1}} & =(p+1) \partial_{[\alpha_{1}} P_{\alpha_{2} \cdots \alpha_{p+1}]} \\
		& =\sum_{j=1}^{p+1}(-1)^{j-1} \partial_{\alpha_{j}} P_{\alpha_{1} \cdots \alpha_{j-1} \alpha_{j+1} \cdots \alpha_{p+1}}.
	\end{aligned}
\end{equation}
The theory has a U(1) gauge symmetry, i.e. the lagrangian is invariant under the transformation $P \to P+d \Lambda$, where $\Lambda$ is an arbitrary $(p-1)$-form field.
The field strength $\mathrm{d}P$ is gauge invariant and satisfies the constraint equation since $\mathrm{d}P$ is closed
\begin{align}
	\mathrm{d}(\mathrm{d}P)\propto \partial_{[\mu} (\mathrm{d} P)_{\alpha_{1} \cdots \alpha_{p+1}]}=0.
\end{align}
The equation of motion of $P$ is obtained to be
\begin{align}
	\nabla_{\alpha_{1}} (Z(r) (\mathrm{d} P)^{\alpha_{1} \cdots \alpha_{p+1}})=\frac{1}{\sqrt{-g}}\partial_{\alpha_{1}}(\sqrt{-g} Z(r) g^{\alpha_{1} \beta_1}\cdots g^{\alpha_{p+1} \beta_{p+1}} (\mathrm{d} P)_{\beta_{1} \cdots \beta_{p+1}})=0.
\end{align}
We still assume that the perturbation propagates along the $x$ direction, $k^\mu=(\omega,q,\cdots)$.
The form field $P$ can be Fourier expanded as 
\begin{align}
	P_{\alpha_{1} \cdots \alpha_{p}}(v,r,x,\boldsymbol{j}) =  \int \frac{\mathrm{d}\omega \mathrm{d}q}{(2\pi)^2} P_{\alpha_{1} \cdots \alpha_{p}}(r,\omega,q) \, e^{-i\omega v+iqx}.
\end{align}
In momentum space, the components of the field $P$ and their corresponding field strength $\mathrm{d}P$ can be divided into three kinds of modes as follows
\begin{small}
	\begin{equation}\label{classification_p}
		\begin{aligned}
			&(1)\text{Pure gauge:}  &P_{vr j_1\cdots j_{p-2}}, P_{vx j_1\cdots j_{p-2}}, P_{rx j_1\cdots j_{p-2}}; \quad &(\mathrm{d} P)_{vrx j_1\cdots j_{p-2}} \\
			&(2)\text{Longitudinal:}  &P_{vj_1\cdots j_{p-1}}, P_{rj_1\cdots j_{p-1}}, P_{xj_1\cdots j_{p-1}}; \quad &(\mathrm{d} P)_{vr j_1\cdots j_{p-1}}, (\mathrm{d} P)_{vx j_1\cdots j_{p-1}}, (\mathrm{d} P)_{rx j_1\cdots j_{p-1}} \\
			&(3)\text{Transverse:}  &P_{j_1\cdots j_{p}}; \quad &(\mathrm{d} P)_{v j_1\cdots j_{p}}, (\mathrm{d} P)_{r j_1\cdots j_{p}}, (\mathrm{d} P)_{x j_1\cdots j_{p}}
		\end{aligned}
	\end{equation}
\end{small}%
where $j_i$ denotes the boundary spatial coordinates except $x$.
This mode decomposition applies in the case of $1<p<d$.
If $p=0$, which is the scalar case, only the transverse mode exist.
If $p=1$, we have the transverse and longitudinal mode and the pure gauge does not exist.
If $p=d$, the pure gauge and longitudinal mode exist and the transverse mode disappears. 
If $p=d+1$, we only have pure gauge mode.
For the case of massless $p$-form field, we find that the field strength of the first channel $(\mathrm{d} P)_{vrx j_1\cdots j_{p-2}}$ must vanish according to its equation of motion. 
This suggests that through an appropriate gauge transformation, all components in this mode can be rendered vanish.
This is why we call it ``pure gauge" mode.
The pure gauge mode is not dynamical and thus does not has pole-skipping phenomenon.
However, in the massive case, the gauge invariance is broken and it is no longer a pure gauge and becomes dynamical as shown in section \ref{sec3.1.1}.
For later convenience, we denote:
\begin{equation}\label{notation}
	\begin{matrix}
        (\mathrm{d} P)_{v r x j_1\cdots j_{p-2}} \to F_g.&\\
		(\mathrm{d} P)_{vr j_1\cdots j_{p-1}} \to F_{l1},&  (\mathrm{d} P)_{vx j_1\cdots j_{p-1}} \to F_{l2},&  (\mathrm{d} P)_{rx j_1\cdots j_{p-1}} \to F_{l3},& \\
		(\mathrm{d} P)_{v j_1\cdots j_{p}} \to F_{t1},&  (\mathrm{d} P)_{r j_1\cdots j_{p}} \to F_{t2},&  (\mathrm{d} P)_{x j_1\cdots j_{p}} \to F_{t3}.&
	\end{matrix}
\end{equation}
where $g$, $l$ and $t$ stand for the pure gauge, longitudinal and transverse mode respectively.

Pole-skipping of the massless longitudinal and transverse modes has been computed in \cite{Wang:2022xoc} using the same method as in the $1$-form case in \ref{sec2.1}.
We show the results as follows.


\paragraph{Longitudinal mode}

The equations of motion corresponding to the longitudinal mode can be reduced to one single second-order ODE for $F_{l2}$
\begin{equation}\label{master_long_massless_p}
	F_{l2}''+a(r)F_{l2}'+b(r)F_{l2}=0.
\end{equation}
Similar to the equation of motion of a massless $1$-form field, there is one master equation that determines the QNM of the longitudinal mode.
Thus the computation for the corresponding pole-skipping points is similar. 
The leading and subleading pole-skipping points for the longitudinal mode of a massless $p$-form field are
\begin{align}
    &\omega=0,\quad q^2=0, \label{p_ml_long_ps0}\\
    &\omega=-i2\pi T,\quad q^2=\pi T \left( (d-2p)h'(r_0)+\frac{2h(r_0) Z'(r_0)}{Z(r_0)}\right).\label{p_ml_long_ps} 
\end{align}
One can also calculate higher-order pole-skipping points using the method in section \ref{sec2.1}.
The results for higher-order pole-skipping in AdS-Schwarzschild black hole are shown in table \ref{table_ml_long}.

\paragraph{Transverse mode}
The master equation for the transverse mode is:
\begin{equation}\label{master_tran_p}
	F_{t3}''+a(r) F_{t3}'-b(r) F_{t3}=0,
\end{equation}
where the coefficients are $a(r)=\frac{2}{r}+\frac{r^2 f'-2i\omega}{r^2 f}+\frac{(d-2p)h'}{2h}+\frac{Z'}{Z}$ and $b(r)=-\frac{2q^2+i(d-2p)\omega h'}{2r^2 f h}-\frac{i \omega Z'}{r^2 f Z}$.
According to the calculation in section \ref{sec2.1}, the leading pole-skipping point is 
\begin{align}\label{p_ml_trans_ps}
    \omega=-i2\pi T,\quad q^2=-\pi T \left( (d-2p)h'(r_0)+\frac{2h(r_0) Z'(r_0)}{Z(r_0)}\right).
\end{align}
Unlike the longitudinal mode, there is no pole-skipping point of $\omega=0$ for the transverse mode, and the pole-skipping starts from $\omega=-i 2\pi T$. 
The results for higher-order pole-skipping in AdS-Schwarzschild black hole are shown in table \ref{table_ml_trans}.

\subsection{Metric fluctuations}\label{sec2.3}

We sketch the computation of the pole-skipping of the stress-energy correlator dual to the metric fluctuations in the bulk \cite{Blake:2019otz}. 

Perturbation of the Einstein-Hilbert action with a cosmological constant
\begin{equation}
    S=\int\mathrm{d}^{d+2} x \sqrt{-g}(R-2\Lambda),\quad g_{\mu\nu}\rightarrow g_{\mu\nu}+h_{\mu\nu}
\end{equation}
leads to the equation of motion of the linearised gravitational field $h_{\mu\nu}$
\begin{equation}\label{eom_massless_gravity}
\begin{aligned}
    \delta R_{\mu\nu}-\frac{1}{2}h_{\mu\nu}(R-2\Lambda)-\frac{1}{2}g_{\mu\nu}\delta R=0,
\end{aligned}
\end{equation}
where $\delta R_{\mu\nu}=\frac{1}{2}(\nabla_{\rho} \nabla_{\mu} h_{\nu}^{\rho} + \nabla_{\rho}\nabla_{\nu} h_{\mu}^{\rho} - \nabla^2 h_{\mu\nu} - \nabla_{\mu}\nabla_{\nu} h)$ and $\delta R = -h^{\mu\nu} R_{\mu\nu} + \nabla_{\mu}\nabla_{\nu}h^{\mu\nu} - \nabla^2 h$.
We assume that the field propagates in the $x$ direction and therefore has a plane-wave expansion $h_{\mu\nu}(v,r,x)=e^{-i\omega v+iqx}h_{\mu\nu}(r)$. The field components are thus divided into three modes: (1) the sound mode containing $h_{vv},h_{vr},h_{vx},h_{rr},h_{rx},h_{xx},h_{ii}$, (2) the diffusive mode containing $h_{vi},h_{ri},h_{xi}$, (3) the tensor mode containing $h_{ij}$ for $i\neq j$. Einstein gravity enjoys diffeomorphism invariance $\delta h_{\mu\nu}=\nabla_{\mu}\xi_{\nu}+\nabla_{\nu}\xi_{\mu}$ under coordinate transformation $\delta x_{\mu}=-\xi_{\mu}$, through which gauge invariants can be constructed. 

\paragraph{Diffusive mode}
The diffusive mode equation of motion can be obtained from eq. \eqref{eom_massless_gravity} as \cite{Blake:2019otz}
\begin{equation}
\frac{\mathrm{d}}{\mathrm{d}r}\left[\frac{h^{d/2+1}}{\omega^2h-q^2r^2f}\left(r^2f\psi'-i\omega\psi\right)\right]+\frac{h^{d/2}}{\omega^2h-q^2r^2f}\left(-i\omega h\psi'-q^2\psi\right)=0,
\end{equation}
for the gauge invariant
\begin{equation}
\psi\equiv\frac{1}{h(r)}\left(\omega h_{xy}+qh_{vy}\right).
\end{equation}
With the near-horizon expansion of $\psi(r)=(r-r_0)^{\lambda}(\sum_{i=0}^{\infty}(r-r_0)^i\psi^{(i)}(r_0))$, one can follow the method aforesaid for the $1$-form field to check that the leading and sub-leading order pole-skipping points are at
\begin{equation}\label{diffps}
\begin{aligned}
    \omega &= 0,\quad q = 0,\\
    \omega &= -2i\pi T,\quad q^2 = d\pi T h'(r_0).  
\end{aligned}
\end{equation}

\paragraph{Sound mode}
The sound mode equation of motion can be obtained from eq. \eqref{eom_massless_gravity} as \cite{Blake:2019otz}
\begin{equation}
\begin{aligned}
&\frac{\mathrm{d}}{\mathrm{d}r}\left[\frac{r^d\left(r^2f\psi'-i\omega\psi\right)}{\left(\omega^2-q^2f-\frac{q^2}{2d}rf'(r)\right)^2}\right]+\frac{r^{d-2}}{\left(\omega^2-q^2f-\frac{q^2}{2d}rf'(r)\right)^2}\left(-i\omega r^2\psi'-q^2\psi\right)\\
&-\frac{\left(d-1\right)q^2r^{d+2}{f'(r)}^2}{2d\left(\omega^2-q^2f-\frac{q^2}{2d}rf'(r)\right)^3}\psi=0,
\end{aligned}
\end{equation}
for the gauge invariant
\begin{equation}
\psi\equiv \frac{1}{r^2}\left(2\omega qh_{vx}+\omega^2 h_{xx}+q^2h_{vv}-\frac{\left(\omega^2-q^2f-\frac{1}{2}q^2rf'(r)\right)}{d-1}h_{ii}\right).
\end{equation}
With the near-horizon expansion of $\psi(r)=(r-r_0)^{\lambda}(\sum_{i=0}^{\infty}(r-r_0)^i\psi^{(i)}(r_0))$, one can find the pole-skipping points of the first three order located at
\begin{equation}\label{sounps}
\begin{aligned}
    \omega &= 2i\pi T,\quad q^2 = -2d\pi T r_0,\\
    \omega &= 0,\quad q^2 = 0,\\
    \omega &= -2i\pi T,\quad q^2=\frac{1}{2}\left(d(d+1)-\frac{8\pi T}{r_0}\pm \sqrt{\left(d(d+1)-\frac{8\pi T}{r_0}\right)^2-\frac{16d^2\pi^2 T^2}{r_0^2}}\right)r_0^2
\end{aligned}
\end{equation}

\section{Pole-skipping for massive fields and the Stueckelberg formalism}\label{sec3}

We now calculate the poleskipping for massive gauge fields. The massive case is more complicated in that gauge invariance is broken in the Lagrangian and more degrees of freedom arise.
Fortunately, we can restore gauge invariance by applying the so-called Stueckelberg massive electromagnetism \cite{Belokogne:2015etf,Ruegg:2003ps}.
Another new feature comes in when we consider a massive form field with $2 \le p \le d+1$.
In this case, the ``pure gauge" mode does not vanish and becomes a new degree of freedom in addition to the longitudinal and transverse modes.
It turns out that pole-skipping also occurs in this mode.
In this section, we first introduce the Stueckelberg formalism for the massive forms.
Then we compute the pole-skipping points for all the three independent modes of these massive fields.
We have developed two computing methods that both involve near-horizon analysis.
The one present in this section is more concise and deals directly with the equation of motion of field components. The other method that deals with the equation of motion of gauge invariant variables is left to appendix \ref{appendixA}. 
At last, the Stueckelberg formalism will be used to interpret the doubling phenomenon in the results.

\subsection{Massive $1$-form fields}\label{sec3.1}

The action for a massive gauge field $A_\mu$ is
\begin{align}\label{massiveA}
    S=\int d^{(d+2)}x \sqrt{-g} Z(\Phi)\left( -\frac{1}{4}  F^{\mu \nu}F_{\mu \nu}-\frac{1}{2}m^2 A^\mu A_\mu \right). 
\end{align}
Note that a coupling of dilaton field $Z(\Phi)$ is also included in the action as in \cite{Blake:2019otz}.
This Lagrangian is not invariant under $A_\mu \to A_\mu+\partial_\mu \Lambda$.
The equation of motion is
\begin{align}\label{1_massive_long_eom}
	\nabla_\mu (Z(r) g^{\mu \alpha}g^{\rho \beta}(\partial_\alpha A_\beta-\partial_\beta A_\alpha))&=m^2 Z(r) g^{\rho \lambda}A_\lambda,
\end{align}
where the equations for $\rho=v, r, x$ correspond to the longitudinal modes, while other $d-1$ equations for $\rho=j$ are all the same and correspond to the transverse modes.

\subsubsection{Longitudinal mode}\label{sec3.1.2}

Let us first deal with the longitudinal mode.
Observe that the equation for $\rho=r$ does not contain any derivatives of $A_r$, we could represent $A_r$ by $A_v$ and $A_x$ which yield
\begin{align}
    A_r=\frac{(q^2+m^2 h(r))A_v+q\omega A_x+i(\omega h(r)A_v'+q r^2 f(r)A_x')}{\omega^2 h(r)-r^2 f(r)(q^2+m^2 h(r))}.
\end{align}
Put this into the equations for $\rho=v$ and $\rho=x$ and we obtain two second-order ODEs which can be expressed in a compact form as follows:
\begin{align}\label{1_massive_long_eom2}
	\psi''+A(r)\psi'+B(r)\psi=0,
\end{align}
where we have combined the two variables into a vector $\psi=(A_v, A_x)^T$. $A(r), B(r)$ are both $2\times 2$ dimensional matrices.\footnote{The full expressions for $A(r)$ and $B(r)$ are complicated. We omit them here and focus only on the leading terms of the $r-r_0$ expansion. Note also that the coefficient matrix $A(r)$ should not be confused with the $1$-form field $A_\mu$.}
We note that $r=r_0$ is the first order pole of both $A(r)$ and $B(r)$.
Their power series expansions are
\begin{equation}\label{ABexp}
	\begin{aligned}
		A(r)=\frac{A_0}{r-r_0}+A_1+A_2 (r-r_0)+ O(r-r_0)^2,\\
		B(r)=\frac{B_0}{r-r_0}+B_1+B_2 (r-r_0)+ O(r-r_0)^2.
	\end{aligned}
\end{equation}
Let the series expansion of an ingoing solution $\psi$ be
\begin{equation}\label{psiexp}
	\begin{aligned}
		\psi=\psi_{0}+\psi_{1} (r-r_0)+ O(r-r_0)^2,
	\end{aligned}
\end{equation}
where $\psi_0=(A^{(0)}_v, A^{(0)}_x) \ne 0$, $\psi_1=(A^{(1)}_v, A^{(1)}_x)$, and put the series expansion \eqref{ABexp} and \eqref{psiexp} into the equation of motion \eqref{1_massive_long_eom2}, it is found that the leading order equation is
\begin{equation}\label{1_massive_long_2}
		A_{0}\psi_{1}+B_0 \psi_0=0.
\end{equation}
If the matrix $A_0$ is invertible, one could solve $\psi_{1}$ once the leading order coefficient $\psi_0$ is given, $\psi_{1}=-A_0^{-1} B_0 \psi_0$。
Then, in general, one could solve order by order to get all higher order coefficients $\psi_n$.
For arbitrary $n$, $\psi_n$ is completely determined by $\psi_0$.
Therefore, the two components $A^{(0)}_v$, $A^{(0)}_x$ of $\psi_0$ are two free parameters of an ingoing solution.
However, $A_0$ is not invertible if $\omega_*$ takes some special value.
If in addition, $q$ also takes an appropriate value $q_*$ correspondingly so that B0 vanishes, there will be three free parameters in an ingoing solution which implies that an extra independent solution appears, and thus the special value ($\omega*$, $q_*$) is a pole-skipping point.

Then we could calculate the leading and subleading order pole-skipping points.
Because $\omega$ appears in the denominator of some terms in the series expansion $A(r)$ and $B(r)$, we discuss the two cases $\omega=0$ and $\omega \ne0$ separately.

(i) When $\omega=0$, the leading coefficient matrices $A_0$ and $B_0$ are
\begin{equation}
	\begin{aligned}
		&A_0=\begin{pmatrix}
			0& 0\\
			0& 1
		\end{pmatrix}, \\
		&B_0=\begin{pmatrix}
			-\frac{(q^2+m^2 h(r_0))}{4 \pi T h(r_0)}& 0\\
			0& -\frac{(q^2+m^2 h(r_0))}{4 \pi T h(r_0)}
		\end{pmatrix}.
	\end{aligned}
\end{equation}
It is easy to see from  
equation \eqref{1_massive_long_2} that if $q^2 \ne -m^2 h(r_0)$, then we must have $A^{(0)}_v=0$ and there are two free parameters for a general ingoing solution: $A^{(0)}_x$ and $A^{(1)}_v$.
Else If $q^2 = -m^2 h(r_0)$, the number of free parameter is three: $A^{(0)}_v$, $A^{(0)}_x$ and $A^{(1)}_v$。
According to the analysis above, the leading pole-skipping point of the longitudinal mode of massive $1$-form field is $(\omega=0,q^2=-m^2 h(r_0))$。

(ii) When $\omega \ne 0$, the leading order coefficient matrices are
\begin{small}
\begin{equation}
	\begin{aligned}
		&A_0=\begin{pmatrix}
			1-\frac{i\omega}{2\pi T}& 0\\
			0& 1-\frac{i\omega}{2\pi T}
		\end{pmatrix}, \\
		&B_0=\begin{pmatrix}
			\frac{2(4\pi T-i\omega)(q^2+m^2 h(r_0))+d \omega^2 h'(r_0)}{8 i\pi T h(r_0)}-\frac{i \omega Z'(r_0)}{4\pi T Z(r_0)}& -\frac{i q}{h(r_0)}\\
			-\frac{i q h'(r_0)}{4\pi T h(r_0)}& -\frac{2(q^2+m^2 h(r_0))+i (d-2) \omega h'(r_0)}{8 \pi T h(r_0)}-\frac{i \omega Z'(r_0)}{4\pi T Z(r_0)}
		\end{pmatrix}.
	\end{aligned}
\end{equation}
\end{small}%
In the special case of $\omega=-i 2\pi T$, all the components of $A_{0}$ vanishes, $A_{0}=0$.
In this case, $\psi_0$ should obey $B_0 \psi_0=0$.
The non-zero solution of $\psi_0$ exists only when the determinant of $B_{0}$ vanishes, otherwise, $\psi_0=0$ which contradicts the assumption form \eqref{psiexp}($\psi_0 \ne 0$) for an ingoing solution.
Therefore, there are two special conditions for the subleading order pole-skipping:
	\begin{equation}\label{2condition}
		\omega=-i 2\pi T \quad \& \quad \text{det}(B_0)=0.
	\end{equation}
It is straightforward to count the free parameters at the pole-skipping points. The two components of $\psi_1$ are both free parameters for the solution since $\psi_1$ is not determined by $\psi_0$.
In addition, one of the components of $\psi_0$ is also free because of the degenerate constraint $B_0 \psi_0=0$.
Thus there are three free parameters in total, which are in contrast to two free parameters in the case of general $\omega$ and $q^2$.
This indicates that the special values for $(\omega,q^2)$ determined by \eqref{2condition} are indeed pole-skipping points.

It turns out that $q^2$ takes two special values under the condition $\eqref{2condition}$,
\begin{small}
\begin{align}\label{q2+-}
	(q^2)_{\pm}=-\left( m^2 h(r_0)+ \pi T h'(r_0) \pm \pi T\sqrt{\left[(d-1)h'(r_0)+2h(r_0) \frac{Z'(r_0)}{Z(r_0)}\right]^2+\frac{4m^2}{\pi T}h(r_0)h'(r_0)}\right).
\end{align}
\end{small}%
This gives two subleading pole-skipping points, located at $\left(-i 2\pi T,(q^2)_+\right)$ and $\left(-i 2\pi T,\right.\\\left.(q^2)_-\right)$, for the longitudinal mode of a massive $1$-form field. 
However, there exists only one subleading pole-skipping point for both the longitudinal and transverse modes of massless $1$-form fields mentioned above.
Moreover, it is also straightforward to check that the situation is the same for higher-order pole-skipping points: the number of pole-skipping points for a massive longitudinal mode is twice as much as for a massless one at every order.
This is the key observation of this paper. As can be seen in the following subsections, similar phenomena exist also in the massive $p$-form fields and massive gravitational fields.

We propose an explanation for this doubling result.
Note that in the massless limit $m\to 0$, we have
\begin{align}
	&q_+^2 \overset{m\to 0}{\rightarrow} \pi T\left[(d-2)h'(r_0)+\frac{2h(r_0) Z'(r_0)}{Z(r_0)}\right],\\
	&q_-^2 \overset{m\to 0}{\rightarrow} -\pi T\left[dh'(r_0)+\frac{2h(r_0) Z'(r_0)}{Z(r_0)}\right].
\end{align}
By comparing pole-skipping results in the massless cases, it is found that $(-i 2\pi T,(q^2)_-)$ becomes the subleading pole-skipping point for a massless scalar field \footnote{The pole-skipping result for a massless scalar field is just \eqref{p_ml_trans_ps} with $p=0$}, while $(-i 2\pi T,q_+^2)$ reduces to the subleading pole-skipping point for the longitudinal mode of a massless $1$-form field.
The higher-order results also show that, in the $m\to 0$ limit, half of the $n$th order pole-skipping points coincide with the pole-skipping points of a massless scalar field and the other half coincide with the pole-skipping points of the longitudinal mode of a massless $1$-form field.

\subsubsection{The Stueckelberg formalism}\label{sec3.1.1}

The reason behind this pole-skipping doubling phenomenon is that a massive vector field has one more degrees of freedom compared with a massless vector field.
We can use Stueckelberg formalism to explain this phenomenon further.

When taking the massless limit in the action of the massive $1$-form field \eqref{massiveA}, the physical degrees of freedom is not continuous in this process.
One of the degree of freedom in the longitudinal mode is lost after $m$ is taken to be $0$.
In order to remove this discontinuity, we apply the Stueckelberg formalism which restores the gauge invariance by introducing an auxiliary scalar field $\phi$. 
The new action is
\begin{align}\label{stuck}
    S=\int d^{(d+2)}x \sqrt{-g} Z(\Phi)\left( -\frac{1}{4}  F^{\mu \nu}F_{\mu \nu}-\frac{1}{2}m^2 (A^\mu+\frac{1}{m}\nabla^\mu \phi) (A_\mu +\frac{1}{m}\nabla_\mu \phi)\right).
\end{align}
It is invariant under the gauge transformation
\begin{align}
    &A_\mu \to A_\mu + \partial_\mu \Lambda,\\
    &\phi \to \phi - m \Lambda,
\end{align}
for an arbitrary scalar function $\Lambda$, so the local U(1) gauge symmetry remains unbroken for the gauge field $A_\mu$.
In addition to the field strength $F_{\mu \nu}$, there is an extra gauge invariant variable
\begin{align}
    J_\mu=A_\mu +\frac{1}{m}\partial_\mu \phi
\end{align}
in the this formalism.
After the auxiliary field is turned on, the theory is equivalent to the massive $1$-form field, and in the meantime, there is no loss of degree of freedom in the massless limit.

In the case of $m\ne 0$, the longitudinal mode contains two coupled degrees of freedom.
Each degree of freedom corresponds to one of the subleading order pole-skipping points \eqref{q2+-}.
On the other hand, in the $m=0$ limit, the $1$-form field and the auxiliary field decouple in the Stueckelberg action \eqref{stuck}.
More specifically, the original two degrees of freedom in the massive longitudinal mode decouple in the Lagrangian.
One of them becomes a massless scalar and the other reduces to the longitudinal mode of a massless vector.
Our result \eqref{q2+-} implies how the pole-skipping behavior for two degrees of freedom changes by tuning the coupling term $m A_\mu \nabla^\mu \phi$\footnote{Note that only the longitudinal components $A_v, A_r, A_x$ are coupled with the scalar field since $\nabla_j \phi=0$, where $j$ denotes transverse coordinates.}.

Therefore the number of pole-skipping points of massive longitudinal mode is doubled compared to the massless case, and the extra half of pole-skipping points arise from pole-skipping points of a massless scalar field at the massless limit.

\subsubsection{Transverse mode}\label{sec3.1.3}

The equation of motion corresponding to the transverse modes in \eqref{1_massive_long_eom} (the equations for $\rho=j$) can be written as
\begin{equation}\label{master_tran_massive_1}
	A_j''+a(r)A_j'+b(r)A_j=0,
\end{equation}
where $a(r)=\frac{2}{r}+\frac{r^2 f'(r)-2i\omega}{r^2 f(r)}+\frac{(d-2)h'(r)}{2h(r)}+\frac{Z'(r)}{Z(r)}$，$b(r)=-\frac{2(q^2+m^2 h(r))+i(d-2)\omega h'}{2r^2 f h}-\frac{i \omega Z'}{r^2 f Z}$.
The horizon $r=r_0$ is again a canonical singularity of this equation.

The calculation for the special points here is the same as the calculation for the $1$-form fields in section \ref{sec2.1}.
The leading coefficients for the power series expansion of $a(r)$ and $b(r)$ around the horizon are
\begin{equation}
	\begin{aligned}
		a_{0}=&1-\frac{2i\omega}{r_0^2 f'(r_0)}=1-\frac{i\omega}{2\pi T}, \\
		b_{0}=&-\left(\frac{2(q^2+m^2 h(r_0))+i(d-2)\omega h'(r_0)}{2r_0^2 f'(r_0) h(r_0)}+\frac{i \omega Z'(r_0)}{r^2 f'(r_0) Z(r_0)}\right)\\
		=&-\left(\frac{2(q^2+m^2 h(r_0))+i(d-2)\omega h'(r_0)}{8\pi T h(r_0)}+\frac{i \omega Z'(r_0)}{4\pi T Z(r_0)}\right).
	\end{aligned}
\end{equation}
Following the previous calculations, the first-order pole-skipping point is
\begin{equation}
	\left\{\begin{matrix}
		a_0=0\\
		b_0=0
	\end{matrix}\right.\; \longrightarrow \; 
	\left\{\begin{matrix}
		\omega=-i2\pi T\\
		q^2=-\pi T \left( (d-2)h'(r_0)+\frac{2h(r_0) Z'(r_0)}{Z(r_0)}\right)-m^2 h(r_0).
	\end{matrix}\right. 
\end{equation}
We note that the special value for $q^2$ changes by an additional mass term $-m^2 h(r_0)$.
Taking the massless limit, the pole-skipping point trivially reduces to the result for the massless transverse case.
This is what we expected from the Stueckelberg action because there is no coupling between the transverse degree of freedom and the auxiliary scalar.
It is also straightforward to calculate the higher-order pole-skipping point by the method in \ref{sec2.1}.

\subsection{Massive $p$-form fields}\label{sec3.2}

\subsubsection{The Stueckelberg formalism}\label{sec3.2.0}

The Stueckelberg formalism has a direct generalization to a $p$-form field \cite{Bizdadea:1998jp}.
The action for a massive $p$-form field $P_{\alpha_1...\alpha_{p}}$ is
\begin{equation}\label{p_massive_action}
	S[P]=-\frac{1}{2(p+1)}\int\mathrm{d}^{d+2} x \sqrt{-g}Z(\phi)((\mathrm{d}P)^2 + (p+1)m^2 P^2).
\end{equation}
The massive term again breaks the gauge invariance of the Lagrangian with $m=0$.
Nevertheless, the theory can restore gauge invariance by simply introducing an auxiliary $(p-1)$- formal field $\varphi$.
Specifically, we perform the following replacement:
\begin{align}
	P_{\alpha_1...\alpha_{p}} \to P_{\alpha_1...\alpha_{p}}+\frac{p}{m} \nabla_{[\alpha_1}\varphi_{\alpha_2...\alpha_{p}]}.
\end{align}
The new action is
\begin{equation}\label{p-stuck}
	\begin{aligned}
		S=&\int \mathrm{d}^{(d+2)}x \sqrt{-g} Z(\Phi)\left( -\frac{1}{2(p+1)}  (\mathrm{d} P)^{\alpha_1\cdots \alpha_{p+1}}(\mathrm{d} P)_{\alpha_1\cdots\alpha_{p+1}}\right.\\
		&\left.-\frac{1}{2}m^2 (P^{\beta_1\cdots\beta_{p}}+\frac{p}{m}\nabla^{[\beta_1} \varphi^{\beta_2\cdots\beta_{p}]}) (P_{\beta_1\cdots\beta_{p}} +\frac{p}{m}\nabla_{[\beta_1} \varphi_{\beta_2\cdots\beta_{p}]})\right)\\
		=&\int \mathrm{d}^{(d+2)}x \sqrt{-g} Z(\Phi)\left(-\frac{1}{2(p+1)}  (\mathrm{d} P)^{\alpha_1\cdots\alpha_{p+1}}(\mathrm{d} P)_{\alpha_1\cdots\alpha_{p+1}}\right.\\
		&\left.-\frac{1}{2}m^2 P^{\beta_1\cdots\beta_{p}}P_{\beta_1\cdots\beta_{p}}-p m P_{\beta_1\cdots\beta_{p}}\nabla^{[\beta_1} \varphi^{\beta_2\cdots\beta_{p}]}-\frac{p^2}{2}\nabla_{[\beta
_
1} \varphi_{\beta_2\cdots\beta_{p}]}\nabla^{[\beta_1} \varphi^{\beta_2\cdots\beta_{p}]}\right).
	\end{aligned}
\end{equation}
It is invariant under the new gauge transformation
\begin{align}
	&P \to P + \mathrm{d} \Lambda,\\
	&\varphi \to \varphi - m \Lambda,
\end{align}
in which $\Lambda$ is an arbitrary $(p-1)$-form field.
In the generalized Stueckelberg theory for a $p$-form field, besides the field strength $\mathrm{d} P$, there is a new gauge invariant: $Q\equiv P+\frac{1}{m}\mathrm{d} \varphi$.

As in the case of a massive $1$-form field, we could use the Stueckelberg formalism to explain the doubling phenomenon of pole-skipping points in the longitudinal channel of a massive $p$-form field, as can be seen in \ref{sec3.3}.

The equation of motion deduced from the action \eqref{p_massive_action} is
\begin{equation}\label{eomdP}
	\begin{aligned}
		\nabla_{\alpha_1} (Z(r) (\mathrm{d} P)^{\alpha_1 \cdots \alpha_{p+1}})&=m^2 Z(r) P^{\alpha_2 \cdots \alpha_{p+1}},
	\end{aligned}
\end{equation}
where the field strength $\mathrm{d} P$ is related with the $p$-form field as
\begin{equation}\label{eomQ}
	\begin{aligned}
		(p+1)\partial_{[\alpha_1} P_{\alpha_2 \cdots \alpha_{p+1}]}&=(\mathrm{d} P)_{\alpha_1 \cdots \alpha_{p+1}}.
	\end{aligned}
\end{equation}
The classification of the components of the $p$-form field and the field strength is the same as the massless case.
Thus we use the same notation as in \eqref{classification_p}.
The new feature in the massive $p \ge 2$ case is that the field strength of the pure gauge mode does not vanish anymore and the mode becomes dynamical.
We also need abbreviations for the components of $P$ in addition to the abbreviations for the field strength as in \eqref{notation}.
We denote the components for each mode of a $p$-form field as
\begin{equation}\label{notation_2}
	\begin{matrix}
        P_{r x j_1\cdots j_{p-2}} \to P_{g1}&  P_{v x j_1\cdots j_{p-2}} \to P_{g2},&  P_{v r j_1\cdots j_{p-2}} \to P_{g3};\\
		P_{v j_1\cdots j_{p-1}} \to P_{l1},&  P_{r j_1\cdots j_{p-1}} \to P_{l2},&  P_{x j_1\cdots j_{p-1}} \to P_{l3};\\& P_{j_1\cdots j_{p}} \to P_t,
	\end{matrix}
\end{equation}
where the subscript $z$, $l$ and $t$ represent the pure gauge, longitudinal and transverse mode respectively.
Next, we compute the pole-skipping points of the three modes separately.

\subsubsection{``Pure gauge" mode}\label{sec3.2.1}

In the massive case, the field components $P_{g1}$, $P_{g2}$ and $P_{g3}$ are dynamical and are not pure gauge anymore. 
Nevertheless, we still refer them as ``pure gauge" mode for convenience.
The equations of motion for the ``pure gauge" mode are 
\begin{equation}\label{ms_p_zero_eom}
	\begin{aligned}
		iq g^{xx} F_z- &=m^2 P_{g3},\\
		\partial_r (W(r) g^{xx} F_z)&=m^2 W(r) g^{xx} P_{g1},\\
		i\omega F_z &=m^2 (P_{g2}+g^{rr} P_{g1}),\\
		\end{aligned}
\end{equation}
where we have used the notation defined in eq.\eqref{notation}, $g^{xx}=\frac{1}{h(r)}$，$g^{rr}=r^2 f(r)$ and $W(r)\equiv \sqrt{-g}(g^{jj})^{p-2}Z(r)=h(r)^{\frac{d}{2}-p+2}Z(r)$.
The relation among $F_z$ and $P_{g1}$, $P_{g2}$, $P_{g3}$ is
\begin{equation}\label{ms_p_zero_cons}	
    F_z=-i\omega P_{g1}-\partial_r P_{g2}+iq P_{g3},
\end{equation}
When $m=0$, the gauge invariant $F_z$ should vanish.
however if $m\neq 0$, this mode becomes dynamical and its pole-skipping pattern can be analyzed.

Put equation \eqref{ms_p_zero_cons} into the equations of motion \eqref{ms_p_zero_eom} and one finds that $P_{g1}$ and $P_{g3}$ can be expressed by $P_{g2}$
\begin{align}
P_{g1}=\frac{(q^2+m^2 h(r))P_{g2}+i\omega h(r)P_{g2}'}{\omega^2 h(r)-r^2 f(r)(q^2+m^2 h(r))},\\
P_{g3}=\frac{\omega q P_{g2}+iq r^2 f(r) P_{g2}'}{\omega^2 h(r)-r^2 f(r)(q^2+m^2 h(r))}.
\end{align}
Substituting the above equation into the second equation of \eqref{ms_p_zero_eom}, we obtain the master equation of the ``pure gauge" mode
\begin{align}\label{p_massive_zero_eom}
    P_{g2}''+a(r)P_{g2}'+b(r)P_{g2}=0,
\end{align}
which has the same form as equation \eqref{1_massive_long_eom2} for the $1$-form field.
By a similar analysis and computation, we find the first and second order pole-skipping points for the ``pure gauge" mode of the massive $p$-form field are 
\begin{equation}\label{ms_zero}
    \begin{aligned}
    &\omega=0,\quad q^2=-m^2 h(r_0),\\
    &\omega=-i2\pi T,\quad q^2=(2+d-2p)\pi T h'(r_0)-h(r_0)\left(m^2-\frac{2\pi T Z'(r_0)}{Z(r_0)}\right).
\end{aligned}
\end{equation}

In the massless limit, the pole-skipping points become
\begin{equation}\label{ml_zero}
    \begin{aligned}
    &\omega=0,\quad q^2=0,\\
    &\omega=-i2\pi T,\quad q^2=\pi T\left((d-2(p-1)) h'(r_0)+\frac{2 h(r_0) Z'(r_0)}{Z(r_0)}\right),
\end{aligned}
\end{equation}
which are identical to the pole-skipping of longitudinal mode of a massless $p-1$-form field (eq.\eqref{p_ml_long_ps0} and eq.\eqref{p_ml_long_ps}).
One can see this from the action \eqref{p-stuck} in which the pure-gauge-related coupling terms can be written as
\begin{align}
	P_{ab\beta_{3}\cdots\beta_{p}}\nabla^{[a} \varphi^{b\beta_3\cdots\beta_{p}]},
\end{align} 
where the indices $a$ and $b$ denote the ``longitudinal coordinates" $v$, $r$ or $x$, while $\beta_j$ denote the ``transverse coordinates" $y_j$.
Note that the component $\varphi^{b\beta_3\cdots\beta_{p}}$ of the Stueckelberg's auxiliary $(p-1)$-form field $\varphi$ belongs to the longitudinal mode. 
Thus the ``pure gauge" mode of a $p$-form field is coupled only with the longitudinal mode of a Stueckelberg $(p-1)$-form field.
Therefore we find that although the massless pure gauge mode is trivial and do not have pole-skipping, the massive ``pure gauge" mode becomes a dynamical degree of freedom and its pole-skipping comes from the Stueckelberg $(p-1)$-form field.

\subsubsection{Longitudinal mode}\label{sec3.2.2}

The equations of motion for the longitudinal mode are
\begin{small}
	\begin{equation}\label{ms_p_long_eom}
		\begin{aligned}
			\partial_r(U(r) g^{rv}g^{vr}F_{l1})-iqU(r)g^{xx}g^{vr}F_{l3}&=m^2 g^{vr} U(r) P_{l2},\\
			-iqU(r)g^{xx}g^{rr}F_{l3}+i\omega U(r)g^{vr}g^{rv}F_{l1}-iqU(r)g^{xx}g^{rv}F_{l2}&=m^2 U(r) (g^{rv} P_{l1}+g^{rr}P_{l2}),\\
			\partial_r(U(r) g^{rv}g^{xx}F_{l2})+\partial_r(U(r) g^{rr}g^{xx}F_{l3})-i\omega U(r)g^{vr}g^{xx}F_{l3}&=m^2 U(r) g^{xx}P_{l3},\\
		\end{aligned}
	\end{equation}
\end{small}%
where we have denoted $g^{rv}=g^{vr}=1$, $g^{xx}=\frac{1}{h(r)}$，$g^{rr}=r^2 f(r)$ and $U(r)\equiv \sqrt{-g}(g^{jj})^{p-1}Z(r)\\=h(r)^{\frac{d}{2}-p+1}Z(r)$.
The relation among $F_{l1}$, $F_{l2}$, $F_{l3}$ and $P_{l1}$, $P_{l2}$, $P_{l3}$ is
\begin{equation}\label{ms_p_long_cons}
	\begin{aligned}
		F_{l1}=-i\omega P_{l2}-\partial_r P_{l1},\\
		F_{l2}=-i\omega P_{l3}-iq P_{l1},\\
		F_{l3}=\partial_r P_{l3}-iq P_{l2}.
	\end{aligned}
\end{equation}
Put equation \eqref{ms_p_long_cons} into the second equation of \eqref{ms_p_long_eom} and one can obtain the solution of $P_{l2}$:
\begin{align}
	P_{l2}=\frac{(q^2+m^2 h(r))P_{l1}+q\omega P_{l3}+i(\omega h(r)P_{l1}'+q r^2 f(r)P_{l3}')}{\omega^2 h(r)-r^2 f(r)(q^2+m^2 h(r))}.
\end{align}
Substituting the above equation into the first and third equation of \eqref{ms_p_long_eom}, we obtain a set of second-order ODE
\begin{align}\label{p_massive_long_eom}
	\psi''+A(r)\psi'+B(r)\psi=0,
\end{align}
where $\psi=(Ql_1, Ql_3)^T$.
$r=r_0$ is a first-order pole of $A(r)$ and $B(r)$.
The form of this equation is the same as equation \eqref{1_massive_long_eom2} in the $1$-form case.
The only difference is a new parameter, the form number $p$, appears in the equation.
By the same analysis and computation, we find the first and second order pole-skipping point for the longitudinal mode of the massive $p$-form field are 
\begin{align}
    &\omega=0,\quad q^2=-m^2 h(r_0),\label{ms_long1}\\
    &\omega=-i2\pi T,\quad q^2=(q^2)_{\pm},\label{ms_long2}
\end{align}
respectively, where
\begin{small}
	\begin{align}
		(q^2)_{\pm}=-\left( m^2 h(r_0)+ \pi T h'(r_0) \pm \pi T\sqrt{\left[(d-2p+1)h'(r_0)+2h(r_0) \frac{Z'(r_0)}{Z(r_0)}\right]^2+\frac{4m^2}{\pi T}h(r_0)h'(r_0)}\right).
	\end{align}
\end{small}%
For higher-order results, the number of pole-skipping points for the massive case is twice as much as the massless case, see table \ref{table_ms_long}.

Comparing the massless limit of this result:
\begin{align}
	&(q^2)_+ \overset{m\to 0}{\rightarrow} \pi T\left[(d-2p)h'(r_0)+\frac{2h(r_0) Z'(r_0)}{Z(r_0)}\right],\\
	&(q^2)_- \overset{m\to 0}{\rightarrow} -\pi T\left[(d-2(p-1))h'(r_0)+\frac{2h(r_0) Z'(r_0)}{Z(r_0)}\right],
\end{align}
with the result of the massless $p$-form fields, we find a similar degenerating structure of pole-skipping as in the case of the $1$-form field.
This is not surprising since the massive $1$-form field considered in \ref{sec3.1.2} is just a special case of $p=1$.
We find that $(\omega=-i2\pi T,(q^2)_+)$ reduces to the subleading order pole-skipping point of the longitudinal mode of massless $p$-form field (i.e.\eqref{p_ml_long_ps}),
while the other special point $(\omega=-i2\pi T,(q^2)_-)$, in the massless limit, turns out to be the leading order pole-skipping point of the \emph{transverse} mode of massless \emph{$p-1$}-form field (i.e.\eqref{p_ml_trans_ps}).
Pole-skipping points of higher order also have analogous behavior.
At each order, half of the pole-skipping points become pole-skipping points of $p$-form massless longitudinal mode and the other half reduces to pole-skipping points of \emph{$p-1$}-form massless \emph{transverse} mode.

The Stueckelberg formalism is also behind this phenomenon.
It is easy to see from the coupling term $P_{\beta_1\cdots\beta_{p}}\nabla^{[\beta_1} \varphi^{\beta_2\cdots\beta_{p}]}$ of the action \eqref{p-stuck} that the longitudinal mode of a $p$-form field is coupled only with the transverse mode of a $(p-1)$-form field but not its longitudinal mode:
\begin{align}
	P_{i j_1 \cdots j_{p-1}}\nabla^{[i} \varphi^{j_1 \cdots j_{p-1}]}\propto \sum_{j} (P_{l1}\, \nabla^{v}\varphi_t+ P_{l2}\, \nabla^{r}\varphi_t + P_{l3}\, \nabla^{x}\varphi_t),
\end{align}
where $i=v,r,x$，$Pl_i \equiv P_{i j_1 \cdots j_{p-1}}$, $\varphi_t$ denotes the transverse mode of the auxiliary $(p-1)$-form field $\varphi_t \equiv \varphi^{j_1 \cdots j_{p-1}}$.
The coupling between $(p-1)$-form $\varphi_t$ and $p$-form $Pl_i$ changes their pole-skipping points.
After the massless limit is taken, the $(p-1)$-form $\varphi_t$ and $p$-form $Pl_i$ decouple in the Lagrangian and thus the pole-skipping points revert to the value of their respective massless cases.

\subsubsection{Transverse mode}\label{sec3.2.3}

The equation of motion for the transverse mode of a massive $p$-form is
\begin{equation}
	\begin{aligned}\label{ms_p_trans_eom}
		\partial_r (V(r) g^{rv} F_{t1}) -i \omega g^{vr} V(r) F_{t2} + \partial_r (V(r) g^{rr} F_{t2}) + i q V(r) g^{xx} F_{t3}=m^2 V(r) P_t,
	\end{aligned}
\end{equation}
where $g^{rv}=g^{vr}=1$, $g^{xx}=\frac{1}{h(r)}$, $g^{rr}=r^2 f(r)$, $V(r)\equiv \sqrt{-g}g^{jj}Z(r)=h(r)^{(\frac{d}{2}-p)}Z(r)$.
According to the relation \eqref{eomQ}, $F_{t1}$, $F_{t2}$, $F_{t3}$, can be expressed by $P_t$ as
\begin{equation}\label{ms_p_trans_cons}
	\begin{aligned}
		F_{t1}=-i\omega P_t,\\
		F_{t2}=\partial_r P_t,\\
		F_{t3}=iq P_t.
	\end{aligned}
\end{equation}
Put the above relation into equation \eqref{ms_p_trans_eom} and we find the master equation for the massive transverse mode:
\begin{equation}\label{master_tran_massive_p}
	Qt''+a(r)Qt'+b(r)Qt=0,
\end{equation}
where the coefficients are 
\begin{equation}
\begin{aligned}
    &a(r)=\frac{2}{r}+\frac{r^2 f'-2i\omega}{r^2 f}+\frac{(d-2p)h'}{2h}+\frac{Z'}{Z},\\
    &b(r)=-\frac{2(q^2+m^2 h)+i(d-2p)\omega h'}{2r^2 f h}-\frac{i \omega Z'}{r^2 f Z}.
\end{aligned}  
\end{equation}
This master equation has the same form as in the massless $1$-form case.
By using the same method, the first-order pole-skipping point for a massive transverse mode is found to be
\begin{equation}\label{ms_trans}
	\left\{\begin{matrix}
		a_0=0\\
		b_0=0
	\end{matrix}\right.\; \longrightarrow \; 
	\left\{\begin{matrix}
		\omega=-i2\pi T\\
		q^2=-\pi T \left( (d-2p)h'(r_0)+\frac{2h(r_0) Z'(r_0)}{Z(r_0)}\right)-m^2 h(r_0)
	\end{matrix}\right.
\end{equation}
When $m=0$, the pole-skipping point reduces back to the transverse massless case, which is expected since the transverse mode is not coupled with the auxiliary field in the Stueckelberg Lagrangian.
The result of higher-order pole-skipping can be found in table \ref{table_ms_trans}.

\subsection{Massive gravitational fields}\label{sec3.3}

The pole-skipping phenomenon was first detected in Einstein gravity, with the leading order skipped poles of sound mode connected with the characteristic parameters of the out-of-time ordered correlation function (OTOC). We have reviewed this part as the standard massless version of gravity in AdS black holes in section \ref{sec2.3}.

To generalize it to the massive case, we consider in this subsection the dRGT massive gravity \cite{deRham:2010ik,deRham:2010kj}. 
We expect to find the doubled skipped poles arising from the extra degree of freedom brought by the breaking of gauge symmetry (i.e. the diffeomorphism for gravity), as we have verified for the form fields. We also discuss in each channel the range of graviton mass $m$ in which the skipped poles lie in the physical region (i.e. real wave number $k_*$).

To start with, we briefly introduce the framework of the dRGT massive gravity, see \cite{Hinterbichler:2011tt,deRham:2014zqa} and references therein for more details. 
The study of the problem of whether there exists a consistent theory that describes massive gravitons can be traced back to Fierz and Pauli \cite{Fierz:1939ix}.
They proposed the following mass term in the Lagrangian
\begin{equation}\label{F-P}
    L_{FP}=-\frac{1}{4}m^2(h_{\mu\nu}h^{\mu\nu}+A h^2)
\end{equation}
and proved that only when $A=-1$ there is no ghost degree of freedom, which would generate instability to a system\cite{Fierz:1939ix}, in the linear order of perturbation.
However, even at this fine tunning coefficient, it was found later by Boulware and Deser \cite{Boulware:1972yco} that the ghost reappears at non-linear orders.
The ghost instability had remained a problem at the full non-linear level of massive gravity until the advent of the dRGT theory. In \cite{deRham:2010ik,deRham:2010kj}, the theory is proved to be ghost-free in the decoupling limit (which would be demonstrated soon) and up to the quintic order away from the limit.
It is confirmed later in \cite{Hassan:2011hr,Hassan:2011tf} that the theory is free of ghost at full non-linearity using the ADM formulation.

To construct a mass term in addition to the Einstein-Hilbert term, a so-called reference metric $f_{\mu\nu}$ is necessary to be contracted with dynamical metric $g_{\mu\nu}$ so that the action is invariant under general coordinate transformations (GCT) \cite{Hassan:2011vm}.
$f_{\mu\nu}$ is usually taken to be Minkowski as originally proposed in \cite{deRham:2010ik,deRham:2010kj}, but in general, it can be arbitrary, even a degenerate one is allowed, see e.g. \cite{Vegh:2013sk,Cao:2015cti}.
In this subsection, we consider two kinds of reference metrics. 
The first one is an AdS black hole with a planar horizon just as considered in previous sections, which is itself also a solution of the theory.  
Therefore the reference metric is identical to the background metric $\Bar{g}_{\mu\nu}$ in this case.
This choice would be make it easier for us to linearise the action and apply the Stueckelberg formalism to analyze the pole-skipping phenomenon.
A similar setup with pure AdS reference metric and $\Bar{g}_{\mu\nu}=f_{\mu\nu}$ was discussed in \cite{Domokos:2015xka}.
The second one is the degenerate metric first introduced in the context of AdS/CFT in \cite{Vegh:2013sk,Davison:2013jba,Blake:2013bqa} in order to introduce momentum dissipation.
For each reference metric, we first solve the metric equation of motion with the ansatz of an asymptotically AdS black hole, and then calculate the pole-skipping in each background.
We focus on 3+1 dimensions ($t,r,x,y$). The generalization to higher dimensions is straightforward.
The action is described by the Einstein-Hilbert term with a cosmological constant $\Lambda$ and dRGT mass terms $\mathcal{U}_i$, $i=2,3,4$
\begin{equation}
    S=-\frac{1}{2\kappa^2}\int\mathrm{d}^4x\sqrt{-g} \Big( R-2\Lambda+m^2\sum_{i=2}^{4}c_i \mathcal{U}_i(g,f)\Big),
\end{equation}
where $g$ is the dynamical metric, $\Lambda$ is the cosmological constant, $c_i$ are constants.  $\mathcal{U}_i$ are symmetric polynomials of the eigenvalues of the matrix $K^{\mu}_{\nu}=\delta^\mu_\nu-(\sqrt{g^{-1}f})^{\mu}_{\nu}$
\begin{equation}
\begin{aligned}
    \mathcal{U}_1 &= [K],\\
    \mathcal{U}_2 &= [K]^2 - [K^2],\\
    \mathcal{U}_3 &= [K]^3 - 3[K][K^2] + 2[K^3],\\
    \mathcal{U}_4 &= [K]^4 - 6[K^2][K]^2 + 8[K^3][K] + 3[K^2]^2 - 6[K^4],
\end{aligned}    
\end{equation}
where the square root in $K$ denotes the matrix square root, i.e. $\sqrt{A}^{\mu}_{\nu}\sqrt{A}^{\nu}_{\rho} = A^{\mu}_{\rho}$, and the square bracket denotes the trace, i.e. $[A]=\mathrm{Tr}(A)$.

To restore the invariance under diffeomorphism, we also introduce a stueckelberg field in this system.  
We parameterizes $f_{\mu\nu}$ as 
\begin{equation}\label{gravityStuck}
		f_{\mu\nu}(x)= \frac{\partial\phi^a}{\partial x^\mu}\frac{\partial\phi^b}{\partial x^\mu}f_{ab}(\phi),
\end{equation}
where the stueckelberg field $\phi^a$ are regarded as new coordinates of the GCT and $f_{ab}$ is the metric in this coordinates.
The configuration of $\phi^a$ corresponds to the gauge choice. A convenient one is the unitary gauge: $\phi^a(x)=x^\mu \delta^a_\mu$
This is the non-linear Stueckelberg formalism for gravity in which the invariance under GCT is restored.
Under GCT, $\phi^a(x)$ transform as scalars $\phi'^a(x')=\phi^a(x)$, while $f_{ab}$ is a fixed metric that does not change so that $f_{\mu\nu}$ indeed transforms as a tensor.

\subsubsection{The reference metric $f_{\mu\nu}=\Bar{g}_{\mu\nu}$}\label{3.4.1}

In the case of the first reference metric, we have $g_{\mu\nu}=f_{\mu\nu}$ in the unitary gauge\footnote{We note that massive gravity with a non-trivial reference metric can be seen as a consistent decoupling limit of bi-gravity theory\cite{Hassan:2011zd,deRham:2014zqa}.}.
Under metric perturbation $h_{\mu\nu}$, $K$ is expanded as
\begin{equation}\label{Kexp}
    K^\mu_\nu=\frac{1}{2}h^\mu_\nu+\frac{1}{8}(h^2)^\mu_\nu+\mathcal{O}(h)^3.
\end{equation}
Therefore it is enough in this case to focus on the term $\mathcal{U}_2$ which becomes the Fierz-Pauli terms \eqref{F-P} (with $A=-1$) at quadratic order of $h_{\mu\nu}$ in the action\footnote{If $g_{\mu\nu}=f_{\mu\nu}$, $\mathcal{U}_2$ is a kind of higher-order extension to the Fierz-Pauli terms that is free of ghost at all order. In general cases of $g_{\mu\nu}\neq f_{\mu\nu}$, the mass terms do not reduce to the Fierz-Pauli form as explained in \cite{Alberte:2011ah,Gumrukcuoglu:2011zh,Koyama:2011wx}.}, and contributes to linear order terms in the equation of motion of $h_{\mu\nu}$.
$\mathcal{U}_3$ and $\mathcal{U}_4$ correspond to higher order terms and $\mathcal{U}_1$ would lead to a theory that pure AdS is not a vacuum solution to the equation of motion \cite{deRham:2014zqa}. 
Thus we choose $c_2=1$ and $c_1=c_3=c_4=0$ for this case.

We find that the equation of motion for the dynamical metric $g_{\mu\nu}$ is
\begin{equation}\label{eomMG}
    G_{\mu\nu}+\Lambda g_{\mu\nu}+m^2(K_{\mu\nu}-g_{\mu\nu}[K]-\frac{1}{2}g_{\mu\nu}([K]^2-[K^2])+K_{\mu\nu}[K]-(K^2)_{\mu\nu})=0,
\end{equation}
where $G_{\mu\nu}$ is the Einstein tensor and we have used the variation formula $\frac{\delta[\mathrm{Tr}\left(\sqrt{g^{-1}f}\right)]}{\delta g^{\mu\nu}}=\frac{1}{2}(\sqrt{g^{-1}f})_{\mu\nu}$ \cite{Cao:2015cti} and the relation $(\sqrt{g^{-1}f})^\mu_\nu=\delta^\mu_\nu-K^\mu_\nu$.
Note that if we want to find a solution satisfying $g_{\mu\nu}=f_{\mu\nu}$ then $K^\mu_\nu=0$ and what determines $g_{\mu\nu}$ is just the usual Einstein equation with a cosmological constant.
Thus the AdS black hole with a planar horizon is a legal solution to this theory, i.e.
    \begin{equation}\label{gbar}
	\mathrm{d}s_g^2=\frac{1}{r^2}(-f(r)\mathrm{d}t^2+\frac{1}{f(r)}\mathrm{d}r^2+\mathrm{d}x^2+\mathrm{d}y^2),
	\end{equation}
where $f(r)=1-\frac{r_0^3}{r^3}$ and AdS radius is set to be $1$.
We denote this metric as $\Bar{g}$ and choose it to be our background solution on which we perturb.

In the following, we analyze the pole-skipping properties of the gravitational sound mode and diffusive mode in the unitary gauge.  
The equation of motion for the gravitational perturbation $h_{\mu\nu}=g_{\mu\nu}-\Bar{g}_{\mu\nu}$ is obtained by an expansion to \eqref{eomMG} about \eqref{gbar}.
Expanding \eqref{eomMG} to linear order of $h_{\mu\nu}$ by using \eqref{Kexp}, the equation of motion is found to be
\begin{equation}\label{linearMG}
    \hat{\mathcal{E}}_{\mu \nu}^{\alpha \beta} h_{\alpha \beta}+\frac{1}{2}m^2(h_{\mu\nu}-\Bar{g}_{\mu\nu}h)=0
\end{equation}
where $\hat{\mathcal{E}}_{\mu \nu}^{\alpha \beta} h_{\alpha \beta}=-\frac{1}{2}\left(\square h_{\mu \nu}-2 \nabla_{\alpha} \nabla_{(\mu} h_{\nu)}^{\alpha}+\nabla_{\mu} \nabla_{\nu} h-\Bar{g}_{\mu \nu}\left(\square h-\nabla_{\alpha} \nabla_{\beta} h^{\alpha \beta}\right)\right)-\frac{\Bar{R}}{4}(h_{\mu\nu}-\frac{1}{2}\Bar{g}_{\mu\nu}h)$\footnote{In comparison with \eqref{eomMG}, we have used the relation $\Lambda=\frac{1}{4}\Bar{R}$ to eliminate the cosmological constant $\Lambda$.}, $h$ is the trace of $h_{\mu\nu}$, $\Bar{R}$ is the scalar curvature of $\Bar{g}$, the covariant derivatives are with respect to $\Bar{g}$ and upper indices are raised by $\Bar{g}^{\mu\nu}$. 
We assume the same plane-wave expansion $h_{\mu\nu}(r,v,x) = e^{i(qx-\omega v)} h_{\mu\nu}(r,\omega,q)$ as used in the massless theory.
With the additional mass term, the linearised gauge invariance $h_{\mu\nu}\to h_{\mu\nu}+\partial_{(\mu}\epsilon_{\nu)}$ is broken and as a consequence more degrees of freedom appear.
Thus one can not combine the components of $h_{\mu\nu}$ into two gauge invariants respectively for the diffusive and sound mode and there is more than one equations of motion for each of the two mode. 
What remains true is that the two modes do not mix each other so that they can still be analyzed separately.

In the following, we analyse the pole-skipping in the Finkelstein-Eddington coordinates $(v,r,x,y)$ in which the metric is \eqref{EFmetric} with $f(r)=1-\frac{r_0^3}{r^3}$ and $h(r)=r^2$.

\paragraph{Diffusive mode}

For diffusive mode, the equations of motion are
\begin{equation}
\begin{aligned}
    \frac{r_0^3-r^3}{2r} h''_{vy}+\frac{i\omega}{2}h'_{vy}+\frac{i(r_0^3-r^3)}{2r}h'_{ry}+\left(\frac{q^2-2i\omega r}{2r^2}+1-\frac{r_0^3}{r^3}-m^2\right)h_{vy}\\
    -i\omega r\left(1-\frac{r_0^3}{r^3}-\frac{i\omega}{2r}\right)h_{ry}+\frac{\omega q}{2r^2}h_{xy}=0\\
    h''_{vy}+i\omega h'_{ry}+\frac{iq}{r^2} h'_{xy}-\frac{2}{r^2}h_{vy}+\left(\frac{q^2+2i\omega r}{r^2}-2m^2\right)h_{ry}-\frac{2iq}{r^3}h_{xy}=0\\
    \frac{r_0^3-r^3}{2r} h''_{xy}+\frac{iq}{2} h'_{vy}+\frac{iq(r^3-r_0^3)}{2r} h'_{ry}+\left(i\omega-\frac{3r_0^3}{2r^2}\right)h'_{xy}+iqr\left(1+\frac{r_0^3}{2r^3}-\frac{i\omega}{2r}\right)h_{ry}\\+\left(1+\frac{2r_0^3}{r^3}-\frac{i\omega}{r}-m^2\right)h_{xy}=0
\end{aligned}
\end{equation}
We discuss the two cases of pole-skipping ($\omega\neq 0$ and $\omega=0$) respectively in the following.

In the case of $\omega\neq 0$, since there is no $h''_{ry}$ in the equations, it is possible to cancel out $h_{ry}$ by solving $h_{ry}$ and $h'_{ry}$ from the first two equations and then substitute the solution (represented by $h_{vy}$, $h_{xy}$ and their first order derivatives) into the third one.
The resulting two independent equations of $h_{vy}$, $h_{xy}$ have the form of 
\begin{equation}\label{diffusiveEOM}
   \psi''+A(r)\psi'+B(r)\psi=0
\end{equation}
where $\psi=(h_{vy}, h_{xy})^T$, $A(r)$ and $B(r)$ are both 2 dimensional coefficient matrices parameterized by $\omega$, $q$, $m$ and $r_0$.
This form of equation appears previously in the case of massive longitudinal mode.
Following the same route, with the leading order coefficients of $A(r)$ and $B(r)$ in the expansion around the horizon
\begin{equation}
	\begin{aligned}
		&A_0=\begin{pmatrix}
			1-\frac{i\omega}{2\pi T}& 0\\
			0& 1-\frac{i\omega}{2\pi T}
		\end{pmatrix}, \\
		&B_0=\begin{pmatrix}
			\frac{(\omega+3ir_0)(2m^2r_0^2-k^2)-6\omega r_0^2}{3\omega r_0^3}& -\frac{i q}{r_0^2}\\
			-\frac{2iq}{3r_0^2}& -\frac{q^2+(6-2m^2)r_0^2-2ir_0\omega}{3r_0^3}
		\end{pmatrix},
	\end{aligned}
\end{equation}
the first order pole-skipping point of the diffusive mode is found to be
\begin{equation}   
   \omega=-2i\pi T=-\frac{3}{2}ir_0,\;\;\;\;q^2=\left(-\frac{3}{2}-m^2\pm\frac{1}{2}\sqrt{81+24m^2})\right) r_0^2.
\end{equation}
Higher order pole-skipping points can be obtained by considering higher order expansions of \eqref{diffusiveEOM} with a similar manner.

In the case of $\omega=0$, only $h_{ry}$ can be represented by the other two variables by solving the first two equations.
But it is enough to cancel out $h_{ry}$ and its derivatives, and the resulting equations are
\begin{equation}
\begin{aligned}
   &h''_{vy}-\left(\frac{q^2+m^2 r^2}{r(r^3-r_0^3)}+\frac{2}{r^2}\right)h_{vy}=0\\
   &h''_{xy}+\left(\frac{3r^2}{r^3-r_0^3}-\frac{q^2+3m^2r^2}{r(q^2+m^2r^2)}\right)h'_{xy}-\left(\frac{q^2+(m^2+6)r^2}{r(r^3-r_0^3)}-\frac{4m^2}{q^2+m^2r^2}\right)h_{xy}=0
\end{aligned}
\end{equation}
The $h_{vy}$ and $h_{xy}$ are decoupled in the above equations. 
We find the equation for $h_{vy}$ has a pole-skipping point at 
\begin{equation}   
   \omega=0,\;\;\;\;q^2=-m^2 r_0^2.
\end{equation}
while the equation for $h_{xy}$ do not has pole-skipping point.

In summary, the highest two order pole-skipping point for the diffusive mode are  
\begin{align}
    &\left(\omega = 0, q^2=-m^2r_0^2\right) \\
    &\left(\omega = -\frac{3}{2}ir_0, q^2 = (-\frac{3}{2}-m^2\pm\frac{1}{2}\sqrt{81+24m^2})r_0^2\right).  
\end{align}
In the $m\to 0$ limit, the above results become
\begin{align}\label{MGdiffusive0}
    &\left(\omega = 0,q^2 = 0\right)  \\
    &\left(\omega = -\frac{3}{2}ir_0,q^2 = 3 r_0^2\right) \;\;\&\;\; \left(\omega = -\frac{3}{2}ir_0, q^2=-6 r_0^2\right).  
\end{align}
There is an extra pole-skipping point at $(\omega= -\frac{3}{2}ir_0,q^2=-6r_0^2)$ compared with result of the massless case (i.e. \eqref{diffps} in section \ref{sec2.3}). 

\paragraph{Sound mode}

For sound mode, it is more convenient to first manipulate the equation of motion \eqref{linearMG} into the following Klein-Gordon-like equation: 
\begin{equation}\label{linearMG2.1}
   (\square-m^2)h_{\nu\rho}+(\Bar{R}_{\mu\nu\rho\sigma}+\Bar{R}_{\mu\rho\nu\sigma})h^{\mu\sigma}=0
\end{equation}
together with two constraint equations:
\begin{equation}\label{linearMG2.2}
   h=0,\;\;\nabla^\mu h_{\mu\nu}=0.
\end{equation}
It is straightforward to verify that \eqref{linearMG2.1} \eqref{linearMG2.2} are equivalent to \eqref{linearMG}\cite{Hinterbichler:2011tt}.
The equations associated with the sound mode are
\begin{equation}\label{linearMGsound}
    \begin{aligned}
     h=0,\;\;\;\;\nabla^\mu h_{\mu a}=0,\\
   (\square-m^2)h_{ab}+(\Bar{R}_{\mu ab \sigma}+\Bar{R}_{\mu ba \sigma})h^{\mu\sigma}=0,\\
   (\square-m^2)h_{yy}+(\Bar{R}_{\mu yy \sigma}+\Bar{R}_{\mu yy \sigma})h^{\mu\sigma}=0,
\end{aligned}
\end{equation}
where $a$ and $b$ denote the coordinates $v$, $r$ or $x$. 
So there are 7 second order differential equations associated with each components of the sound mode, 3 first order and 1 zeroth order differential equations that give 4 constraints and reduce the number of degrees of freedom to 3.
Instead of reducing the number of equations, we put the expansion $h_{\mu\nu}(r)=(r-r_0)^\lambda\sum_{i=0}^{\infty}h_{\mu\nu}^{(i)}(r_h)(r-r_h)^i$ directly into \eqref{linearMGsound} and solving the equations order by order.
At the order of $(r-r_0)^{-1}$, the ingoing boundary condition at the horizon requires the leading exponent $\lambda$ to vanish.
At the order of $(r-r_0)^{i}, (i=0,1,2,...)$, we find the relation among $h_{\mu\nu}^{(n)}, (n=0,1,2,...,i+1)$ can be written as the following matrix equation
\begin{align}\label{MGleomi}
    C_{i+1}\psi_{i+1}+C_{i}\psi_{i}+...+C_0\psi_{0}=0
\end{align}
where we denote $\psi_i=(h_{vv}^{(i)}, h_{vx}^{(i)}, h_{xx}^{(i)}, h_{yy}^{(i)}, h_{vr}^{(i)}, h_{rr}^{(i)}, h_{rx}^{(i)})^T$, $C_{i}, (i=0,1,2,...)$ are $11\times 7$ dimensional matrices parameterized by $\omega$, $q$, $r_0$ and $m$.
As pointed out in section \ref{sec2.1}, pole-skipping occurs at the special values of the pair $(\omega,q)$ at which there is an additional free parameter in general solution (and thus an additional independent ingoing solution).
To count the free parameters, we start from the equation of order $(r-r_0)^{0}$:
\begin{align}\label{MGleom0}
    C_{1}\psi_{1}+C_0\psi_{0}=0.
\end{align}
By linear transformations among the 11 equations in \eqref{MGleom0}, constraints to the components of $\psi_0$ can be obtained.
For general $\omega$ and $q$ (i.e. do not coincide with any pole-skipping point), there are 4 constraints which can be written as the matrix form
\begin{equation}\label{constraint}
    M_0 \psi_{0}=0,
\end{equation}
where $M_0$ is a $4\times 7$ dimensional matrix:
\begin{small}
   \begin{equation}\label{constraintM}
    \begin{pmatrix}
  0&  0&  \frac{1}{r_0^2}&  \frac{1}{r_0^2}&  2&  0& 0\\
  \frac{q^2+m^2r_0^2-2i\omega r_0}{r_0^2(3r_0+2i\omega)}&  -\frac{iq}{r_0^2}&  0&  0&  i\omega&  0& 0\\
  -\frac{q}{r_0 \omega}&  -\frac{i(q^2+m^2 r_0^2-4i\omega r_0)}{2r_0^2 \omega}&  -\frac{iq}{r_0^2}&  0&  0&  0& i\omega-\frac{3r_0}{2}\\
  \frac{-2r_0}{r_0^3(3r_0-2i\omega)}&  \frac{2iq}{r_0^3(3r_0-2i\omega)}&  \frac{2i\omega}{r_0^3(3r_0-2i\omega)}&  \frac{2i\omega}{r_0^3(3r_0-2i\omega)}&  \frac{r_0(q^2+(6+m^2)r_0^2-2i\omega r_0)}{r_0^3(3r_0-2i\omega)}&  i\omega-3r_0& -\frac{iq}{r_0^2}
\end{pmatrix}.
\end{equation}
\end{small}
and $\psi_0=(h_{vv}^{(0)}, h_{vx}^{(0)}, h_{xx}^{(0)}, h_{yy}^{(0)}, h_{vr}^{(0)}, h_{rr}^{(0)}, h_{rx}^{(0)})^T$.
Thus only 3 components of $\psi_0$ are free parameters of the solution, which means that $\psi_1$ can be determined by these free parameters.
The coefficients of higher orders $\psi_i\;(i>1)$ can be solved iteratively from eq.\eqref{MGleomi} and they also depend only on the three free parameters.
But there may be more free parameters at some special values of $\omega$ and $q$. 
For instance, the number of constraint decreases to 3 for $\omega=\frac{3}{2}ir_0$ or $\omega=0$ and appropriate values of $q$.
The decreasing of constraint indicate one more free parameter than general cases.
Moreover, if $\omega=-\frac{3}{2}ir_0$, 2 components of $\psi_1$ ($h_{xx}^{(1)}$ and $h_{yy}^{(1)}$) also become free.
We discuss these three special value of $\omega$ case by case in the following.

If $\omega=\frac{3}{2}ir_0$, the first constraint in \eqref{constraint}, i.e. the first line of the matrix \eqref{constraintM}, changes to
\begin{equation}\label{constraint1}
    \frac{q^2+(3r_0^2+m^2 r_0^2)}{r_0^2}h_{vv}^{(0)}=0.
\end{equation}
The special value for $q$ is identified as $q^2=-(3+m^2)r_0^2$, since the constraint \eqref{constraint1} disappears at this value of $q$ and there must be one more free parameters.

If $\omega=0$, the constraints at $(r-r_0)^0$ order is 
\begin{equation}\label{constraint2}
    \begin{pmatrix}
  0&  0&  \frac{1}{r_0^2}&  \frac{1}{r_0^2}&  2&  0& 0\\
  \frac{q^2+m^2r_0^2}{3r_0^3}&  -\frac{iq}{r_0^2}&  0&  0&  0&  0& 0\\
  \frac{2iq}{r_0}&  -(m^2+\frac{q^2}{r_0^2})&  0&  0&  0&  0& 0\\
  \frac{2}{3r_0^3}&  \frac{2iq}{3r_0^4}&  \frac{2i\omega}{r_0^3(3r_0-2i\omega)}&  \frac{2i\omega}{r_0^3(3r_0-2i\omega)}&  \frac{q^2+(6+m^2)r_0^2}{3r_0^3}&  3r_0& \frac{iq}{r_0^2}
\end{pmatrix}\psi_0=0.
\end{equation}
The special value for $q$ is obtained by the condition that the second and the third equation of \eqref{constraint2} are proportional, such that there are again one less constraint.
This condition gives
\begin{equation}
    q^2=-(3+m^2\pm\sqrt{9+6m^2})r_0^2
\end{equation}

If $\omega=-\frac{3}{2}ir_0$, 5 independent constraints at $(r-r_0)^0$ order are present:
\begin{equation}\label{constraint3}
    \begin{pmatrix}
  0&  0&  \frac{1}{r_0^2}&  \frac{1}{r_0^2}&  2&  0& 0\\
  \frac{q^2+(m^2-3)r_0^2}{6r_0^3}&  -\frac{iq}{r_0^2}&  0&  0&  \frac{3r_0}{2}&  0& 0\\
  -\frac{2iq}{3r_0^2}&  \frac{q^2+(m^2-6)r_0^2}{3r_0^3}&  -\frac{iq}{r_0^2}&  0&  0&  0& 0\\
  2&  \frac{4iq}{r_0}&  -\left(3+m^2+\frac{q^2}{r_0^2}\right)&  0&  -6r_0^2&  0& 0\\
  2&  0&  0&  -\left(3+m^2+\frac{q^2}{r_0^2}\right)&  -6r_0^2&  0& 0
\end{pmatrix}\psi_0=0,
\end{equation}
Equation \eqref{constraint3} force $h_{vv}^{(0)}$, $h_{vx}^{(0)}$, $h_{xx}^{(0)}$, $h_{yy}^{(0)}$, $h_{vr}^{(0)}$ to vanish, while $h_{rr}^{(0)}$, $h_{rx}^{(0)}$ are free.
Besides the 2 free parameters in $\psi_0$, there are also 2 free parameters coming from $\psi_1$, $h_{xx}^{(1)}$ and $h_{yy}^{(1)}$, unlike the previous two cases where every component of $\psi_1$ can be represented by a linear combination of components of $\psi_0$.
It seems there are 4 free parameters in total, but it turns out that there is another constraint at order $(r-r_0)^1$,
\begin{equation}
   \frac{1}{r_0^2}h_{xx}^{(1)}+\frac{1}{r_0^2}h_{yy}^{(1)}-3r_0 h_{rr}^{(0)}-\frac{2iq}{r_0^2}h_{rx}^{(0)}=0
\end{equation}
that reduce the number to be 3, the same number as in general case.
Hence, the special points of pole-skipping must come from the condition that the 5 equations in \eqref{constraint3} are linear dependent, since the linear dependency give us 1 more free parameter in $h_{vv}^{(0)}$, $h_{vx}^{(0)}$, $h_{xx}^{(0)}$, $h_{yy}^{(0)}$, $h_{vr}^{(0)}$ instead all vanishing.
We find that, in order to fulfill the condition, $q^2$ should obey the polynomial equation
\begin{equation}
\begin{aligned}
       q^8+(15+4m^2)r_0^2 q^6+3(9+7m^2+2m^4)r_0^4 q^4+(135+18m^2-3m^4+4m^6)r_0^2 q^2\\
       +(162+135m^2-9m^4-9m^6+m^8)r_0^8=0.
\end{aligned}
\end{equation}
We have also checked that the number of free parameters can not be reduced by higher order equations.

In summary, the highest three order of pole-skipping point for the sound mode are (or determined by) 
\begin{equation}
    \begin{aligned}
    &\left(\omega = \frac{3}{2}ir_0, q^2 = -(3+m^2)r_0^2\right)  \\
    &\left(\omega = 0, q^2  = -(3+m^2\pm\sqrt{9+6m^2})r_0^2\right)\\
    &\left(\omega = -\frac{3}{2}ir_0, q^8+(15+4m^2)r_0^2 q^6+3(9+7m^2+2m^4)r_0^4 q^4\right.\\
    &\left.+(135+18m^2-3m^4+4m^6)r_0^2 q^2+(162+135m^2-9m^4-9m^6+m^8)r_0^8=0\right).
\end{aligned}
\end{equation}
In general, we find that pole-skipping points for the massive gravitational sound mode locate at $\omega=-2ni\pi T=-\frac{3}{2}nir_0$ with $n=-1,0,1,2,3,...$ and corresponding values of $q$.
There are more pole-skipping with $n>1$ which we choose not show here but can be obtained using our method order by order.
In the $m\to 0$ limit, the above results become
\begin{equation}\label{MGsound0}
    \begin{aligned}\
    &\left(\omega = \frac{3}{2}ir_0, q^2 = -3 r_0^2\right)  \\
    &\left(\omega = 0, q^2 = 0\right) \;\;\&\;\; \left(\omega = 0, q^2=-6r_0^2\right)\\
    &\left(\omega = -\frac{3}{2}ir_0, q^2 = -\frac{3}{2}(5\pm \sqrt{17})r_0^2\right) \;\;\&\;\; \left(\omega = -\frac{3}{2}ir_0, q^2=\pm 3ir_0^2\right)
\end{aligned}
\end{equation}
The pole-skipping at $(\omega=0,q^2=-6r_0^2)$ and $(\omega=-\frac{3}{2}ir_0,q^2=-\frac{3}{2}(5\pm \sqrt{17})r_0^2)$ are newly appeared compared with the result of massless case (i.e. eq.\eqref{sounps} in section \ref{sec2.3}). 

\paragraph{Stueckelberg}
As the next step, in order to test if the Stueckelberg formalism could explain the doubling of pole-skipping, we turn on perturbations about the unitary gauge, $\phi^a=x^\mu \delta^a_\mu-A^a$, where $A^a$ are linearized Stueckelberg fields.
According to eq.\eqref{gravityStuck}, $f_{\mu\nu}$ changes as
\begin{equation}
f_{\mu\nu}\to f_{\mu\nu}-2\nabla_{(\mu} A_{\nu)}+\mathcal{O}(A^2)
\end{equation}
In comparison with \eqref{Kexp}, the matrix $K^{\mu}_{\nu}$ now changes to 
\begin{equation}
\begin{aligned}
    K^{\mu}_{\nu}=&\delta^\mu_\nu-\left(\sqrt{g^{-1}f}\right)^{\mu}_{\nu}\\
    =&\delta^\mu_\nu-\sqrt{(\Bar{g}^{\mu\rho}-h^{\mu\rho})(f_{\rho\nu}-2\nabla_{(\rho} A_{\nu)})}\\
    =&\delta^\mu_\nu-\sqrt{\delta^\mu_\nu-h^\mu_\nu-2\Bar{g}^{\mu\rho}\nabla_{(\rho} A_{\nu)}}\\
    =&\frac{1}{2}\Bar{g}^{\mu\rho}\left(h_{\rho\nu}+2\nabla_{(\rho} A_{\nu)}\right),
\end{aligned}
\end{equation}
where quadratic and higher order terms of $h$ and $A$ are omitted and $f=\Bar{g}$ is used in the third line. 
It follows that the metric fluctuation $h_{\mu\nu}$ changes effectively as
\begin{equation}
h_{\mu\nu}\to h_{\mu\nu}+2\nabla_{(\mu} A_{\nu)}
\end{equation} 
such that one should make this substitution in the $h_{\mu\nu}$'s equation of motion \eqref{linearMG}.
Then the action for the fluctuations becomes (up to the quadratic order)
\begin{equation}
    \begin{aligned}
S=\int \mathrm{d}^{4} x \; \mathcal{L}_{m=0}+\sqrt{-\Bar{g}}&\left[-\frac{1}{4} m^{2}\left(h_{\mu \nu} h^{\mu \nu}-h^{2}\right)-\frac{1}{4} m^{2} F_{\mu \nu} F^{\mu \nu}+\frac{1}{4} m^{2} \Bar{R} A^{\mu} A_{\mu}\right.\\
&\left.- m^{2}\left(h_{\mu \nu} \nabla^{\mu} A^{\nu}-h \nabla_{\mu} A^{\mu}\right)\right]
\end{aligned}
\end{equation}
where $\mathcal{L}_{m=0}=\sqrt{-\Bar{g}} h^{\mu\nu}\hat{\mathcal{E}}_{\mu \nu}^{\alpha \beta} h_{\alpha \beta}$ is the kinetic term of a massless graviton, $F_{\mu\nu}=\nabla_{\mu}A_\nu-\nabla_\nu A_\mu$ and we have used the relation $\nabla_{\mu} A_{\nu} \nabla^{\nu} A^{\mu}=\left(\nabla_{\mu} A^{\mu}\right)^{2}-\Bar{R}_{\mu \nu} A^{\mu} A^{\nu}$.
We note that the extra terms (compared with the F-P terms) in the square brackets contain exactly the Lagrangian for a massive vector field with the mass proportion to the background scalar curvature as well as interacting terms between the graviton $h_{\mu\nu}$ and the massive vector $A_{\mu}$. 
To normalize the kinetic term of the vector field such that we have the correct factor $-\frac{1}{4}$ for the kinetic term, we rescale the field as $A_\mu\to \hat{A}_\mu=\frac{A_\mu}{m}$.
Now the action is
\begin{equation}\label{mGRStuck}
    \begin{aligned}
S=\int \mathrm{d}^{4} x \; \mathcal{L}_{m=0}+\sqrt{-\Bar{g}}&\left[-\frac{1}{4} m^{2}\left(h_{\mu \nu} h^{\mu \nu}-h^{2}\right)-\frac{1}{4} \hat{F}_{\mu \nu} \hat{F}^{\mu \nu}+\frac{1}{4} \Bar{R} \hat{A}^{\mu} \hat{A}_{\mu}\right.\\
&\left.- m\left(h_{\mu \nu} \nabla^{\mu} \hat{A}^{\nu}-h \nabla_{\mu} \hat{A}^{\mu}\right)\right]
\end{aligned}
\end{equation}
Then we take the limit $m\to 0$ while keep $\hat{A}^{\mu}$ invariant.
In this limit, the 5 degrees of freedom of massive graviton separate to two decoupled sections, one is a massless graviton propagating 2 degrees of freedom, the other is a massive vector field with 3 degrees of freedom \footnote{This is a known result that there is no vDVZ (van Dam, Veltman, Zakharov \cite{vanDam:1970vg,Zakharov:1970cc}) discontinuity on curved spacetimes \cite{Hinterbichler:2011tt}.} and the mass is identified to be $m_A^2=-\frac{\Bar{R}}{2}=6$ by comparison with the Lagrangian of a massive vector field \eqref{massiveA}.

After turning on the Stueckelberg fields $A_\mu$ and taking the decoupling limit, it is straightforward to match the pole-skipping points in this limit \eqref{MGsound0} with that of the massless graviton $h_{\mu\nu}$ and massive vector $A_\mu$.
By observing the interaction terms in \eqref{mGRStuck} in the case that the diffusive modes of $h_{\mu\nu}$ (i.e. $h_{vy}$, $h_{ry}$ and $h_{xy}$) are turned on and all sound modes (i.e. $h_{vv}$, $h_{vr}$, $h_{vx}$, $h_{rr}$, $h_{rx}$, $h_{xx}$ and $h_{yy}$) vanish, we find 
\begin{equation}
\begin{aligned}
    h_{\mu \nu} \nabla^{\mu} \hat{A}^{\nu}-h \nabla_{\mu} \hat{A}^{\mu}=&h_{ay}\nabla^a \hat{A}^{y}+h_{ya}\nabla^y \hat{A}^{a}\\
    =&h_{ay}(g^{yy}\nabla^a \hat{A}_{y}+g^{yy}g^{ab}\nabla_y \hat{A}_{b})\\
    =&h_{ay}g^{yy}(\nabla^a-g^{ab}\Gamma^{y}_{yb}) \hat{A}_{y}
\end{aligned}
\end{equation}
where the index $a$ denotes $v$, $r$, or $x$.
We have used $h=0$ in the first line of the above equation, and $\partial_y \hat{A}_b=0$ together with $\Gamma^{a}_{yb}=0$ to obtain the third line.
Therefore, the diffusive mode of $h_{\mu\nu}$ are coupled only with the transverse mode of the vector field $\hat{A}_{\mu}$ (i.e. $\hat{A}_{y}$).
The rest components of $h_{\mu\nu}$ (i.e. the sound modes) are coupled with the longitudinal part of $\hat{A}_{\mu}$ (i.e. $\hat{A}_{v}$, $\hat{A}_{r}$ and $\hat{A}_{x}$). 
This suggests that, in the decoupling limit, the extra pole-skipping points of the gravitational diffusive mode should also be the pole-skipping points of the transverse mode of a massive $1$-form field.
Moreover, the extra pole-skipping points of the gravitational sound mode and the longitudinal mode of a massive $1$-form field should also match in the decoupling limit.
This is exactly the case for our results, since the extra pole-skipping point of gravitational diffusive mode $(\omega=-\frac{3}{2}i r_0,q^2=-6r_0^2)$ in \eqref{MGdiffusive0} matches with \eqref{ms_trans} for the transverse mode of a massive $1$-form, and also the extra pole-skipping points of the gravitational sound mode $(\omega=0,q^2=-6r_0^2)$ and $(\omega=-\frac{3}{2}i r_0,q^2=-\frac{3}{2}(5\pm \sqrt{17})r_0^2)$ in \eqref{MGsound0} is just \eqref{ms_long1} and \eqref{ms_long2} respectively for the longitudinal mode of a massive $1$-form.\footnote{To match the pole-skipping points, we assumed $Z=1$ in the gravitational side, and used the relation $m_A=6$.}

It is also straightforward to explain why the extra gravitational skipped poles emerge at sub-leading (and higher) order: classifying the order by the frequency $\omega=-2\pi i(n-l)T$ with $n=1,2,3,...$, we shall find that both the sub-leading order skipped poles of the gravitational sound mode ($n=2, l=2$) and the leading order skipped poles of the vector longitudinal mode ($n=1, l=1$) share the same special value of frequency ($\omega=0$), and that the sub-leading order skipped poles of the gravitational diffusive mode ($n=2, l=1$) and the leading order skipped poles of the vector transverse mode ($n=1, l=0$) also have the same special value of frequency ($\omega=-2\pi i T$).

To summarize, the extra pole-skipping points originate from the additional degree of freedom which correspond to the Stueckelberg fields.


\subsubsection{The reference metric $f_{\mu\nu}\neq \Bar{g}_{\mu\nu}$}\label{3.4.2}

To generalize the case of $f_{\mu\nu}=g_{\mu\nu}$ to be of non-Fierz-Pauli type, we consider in this subsection a type of nonlinear dRGT gravity with broken diffeomorphism dual to broken translational symmetry in the boundary theory \cite{Vegh:2013sk,Davison:2013jba,Blake:2013bqa}, which does not reduce to the FP theory at linear order. With the prescription of \cite{Vegh:2013sk}, in which the reference metric is set to be $f=\text{diag}(0,0,1,1)$, we expect to find the extra skipped poles arising from the extra degrees of freedom brought by the mass term, as we have verified for the gauged matter coupling with form symmetry and for the case of $f_{\mu\nu}=g_{\mu\nu}$ above.

With the reference metric $f$ given, 
one can solve the gravitational field equation of motion of the dynamical metric $g$. 
Flat $f$ can be obtained with a coordinate transformation $\phi^a(x)$ on the flat spacetime metric $\eta$:
\begin{equation}
    f_{\mu\nu} = \partial_{\mu}\phi^a \partial_{\nu}\phi^b \eta_{ab},
\end{equation}
where different $\phi^a$ fields correspond to different gauges. For the unitary gauge, $\phi^a = x^{\mu}\delta^a_{\mu}$. \cite{Vegh:2013sk} focused on a special case of $f_{\mu\nu} = \mathrm{diag}(0,0,1,1)$ such that transformations by arbitrary $\phi$ fields satisfying $\phi^x = x,\;\phi^y = y$ keep the reference metric and consequently the theory invariant. Therefore, in this case the covariance is preserved in $t$ and $r$ directions and is broken in the spatial directions.


Following the convention in \cite{Vegh:2013sk}, we define another matrix $\tilde{K}^{\mu}_{\nu}=\delta^{\mu}_{\nu}-K^{\mu}_{\nu}=(\sqrt{g^{-1}f})^{\mu}_{\nu}$, and the equation of motion of the dynamical metric is find to be:
\begin{equation}
    R_{\mu\nu} - \frac{R}{2}g_{\mu\nu} + \Lambda g_{\mu\nu} + m^2 X_{\mu\nu} = 0,
\end{equation}
where
\begin{equation}\label{Xmunu}
    X_{\mu\nu} = -\frac{\alpha}{2}([\tilde{K}]g_{\mu\nu}-\tilde{K}_{\mu\nu}) - \beta\Big( (\tilde{K}^2)_{\mu\nu} - [\tilde{K}]\tilde{K}_{\mu\nu} + \frac{g_{\mu\nu}}{2}([\tilde{K}]^2 - [\tilde{K}^2]) \Big).
\end{equation}
It is noteworthy that only the two lowest $\mathcal{U}_i$ terms ($i=1,\;2$) are kept here considering the choice of the reference metric where the $c_i$ coefficients are replaced with $\alpha$ and $\beta$ for simplicity.

We use the Finkelstein-Eddington coordinates $(v,r,x,y)$ for the pole-skipping analysis in the following 
\footnote{Note that the $r$-coordinate in \cite{Vegh:2013sk} is the inverse of our $r$-coordinate.}. By using the static symmetric ansatz of the metric, i.e. \eqref{EFmetric} in 3+1 dimensions, we obtain the solution of the metric function $f(r) = 1 + \frac{m^2\alpha}{2r} + \frac{m^2\beta}{r^2} - \frac{r_0^3}{r^3}$, and the horizon radius $r_h$ is determined by $f(r_h) = 0$. Thus the geometry has a background temperature at $T = \frac{3r_h}{4\pi}+\frac{m^2}{4\pi}(\alpha+\frac{\beta}{r_h})$.

In a similar manner, we analyze the pole-skipping properties of the gravitational sound mode and diffusive mode in the two following subsections. For the gravitational perturbation, we assume the same plane-wave expansion $h_{\mu\nu}(r,v,x) = e^{i(qx-\omega v)} h_{\mu\nu}(r,\omega,q)$ as used in the massless theory, and therefore perturb the dynamical metric as $\delta g_{\mu\nu} = h_{\mu\nu}$. We calculate $\delta \tilde{K}_{\mu\nu}$ and $\delta X_{\mu\nu}$ in eq. \eqref{Xmunu}:
\begin{equation}
\begin{aligned}
    \delta (\tilde{K}^2)^{\mu}_{\nu} &= \delta (g^{\mu\rho}\tilde{K}_{\rho\theta}g^{\theta\lambda}\tilde{K}_{\lambda\nu}) \\
    &= -h^{\mu\rho}\tilde{K}_{\rho\theta}g^{\theta\lambda}\tilde{K}_{\lambda\nu} + g^{\mu\rho}\delta \tilde{K}_{\rho\theta}g^{\theta\lambda}\tilde{K}_{\lambda\nu} - g^{\mu\rho}\tilde{K}_{\rho\theta}h^{\theta\lambda}\tilde{K}_{\lambda\nu} + g^{\mu\rho}\tilde{K}_{\rho\theta}g^{\theta\lambda}\delta \tilde{K}_{\lambda\nu},
\end{aligned}
\end{equation}
and combine with $\delta (\tilde{K}^2)^{\mu}_{\nu} = -h^{\mu\lambda}f_{\lambda\nu}$, and insert the solution of $\delta \tilde{K}_{\mu\nu}$ into $\delta X_{\mu\nu}$. 

Specifically, we evaluate the non-zero components of $\delta \tilde{K}_{\mu\nu}$ to be
\begin{equation}
\begin{pmatrix}
\delta \tilde{K}_{xx} & \delta \tilde{K}_{xy} \\
\delta \tilde{K}_{yx} & \delta \tilde{K}_{yy}
\end{pmatrix}=e^{i(qx-\omega v)}
\begin{pmatrix}
\frac{h_{xx}(r)}{2r} & \frac{h_{xy}(r)}{2r} \\
\frac{h_{xy}(r)}{2r} & \frac{h_{yy}(r)}{2r}
\end{pmatrix}.
\end{equation}
Insert $\delta \tilde{K}_{\mu\nu}$ and we evaluate $\delta X_{\mu\nu}$ to be
\begin{equation}
\delta X_{\mu\nu}=e^{i(qx-\omega v)}
\begin{pmatrix}
\delta X_{vv} & \delta X_{vr} & \frac{(r\alpha-\beta)h_{vx}(r)}{r^2} & \frac{(r\alpha-\beta)h_{vy}(r)}{r^2}\\
\delta X_{rv} & \frac{(r\alpha-\beta)h_{rr}(r)}{r^2} & \frac{(r\alpha-\beta)h_{rx}(r)}{r^2} & \frac{(r\alpha-\beta)h_{ry}(r)}{r^2} \\
\frac{(r\alpha-\beta)h_{vx}(r)}{r^2} & \frac{(r\alpha-\beta)h_{rx}(r)}{r^2} & \frac{\alpha(2h_{xx}(r)-h_{yy}(r))}{4r} & \frac{3\alpha h_{xy}(r)}{4r} \\
\frac{(r\alpha-\beta)h_{vy}(r)}{r^2} & \frac{(r\alpha-\beta)h_{ry}(r)}{r^2} & \frac{3\alpha h_{xy}(r)}{4r} & \frac{\alpha(2h_{yy}(r)-h_{xx}(r))}{4r}
\end{pmatrix},
\end{equation}
where
\begin{small}
   \begin{equation*}
\begin{aligned}
    \delta X_{vv} &= \frac{(r\alpha-\beta)h_{vv}(r)}{r^2}\\
    &+\frac{(2r^4\alpha+4r_0^3\beta+4m^2r^2\alpha\beta-r^3(m^2\alpha^2+4\beta)-2r(r_0^3\alpha+2m^2\beta^2))(h_{xx}(r)+h_{yy}(r))}{8r^5}\\
    \delta X_{vr} &= \delta X_{rv} = \frac{(r\alpha-\beta)h_{vr}(r)}{r^2}-\frac{(r\alpha-2\beta)(h_{xx}(r)+h_{yy}(r))}{4r^4}.
\end{aligned}
\end{equation*} 
\end{small}

\paragraph{Sound mode}

With the pertubations $\delta X_{\mu\nu}$, we obtain a set of sound mode equations of motion, for which we only show the $vv$-component for brevity:
\begin{equation}\notag
\begin{aligned}
    &2r(-2r^3+2r_0^3+m^2r^2\alpha-2m^2r\beta)^2(q^2r+10r^3+2r_0^3+2m^2r\beta-r^2(3m^2\alpha+2i\omega))h_{rr}(r)\\
    &-2iqr(-2r^3+2r_0^3+m^2r^2\alpha-2m^2r\beta)(8r^3-2r_0^3+4m^2r\beta-r^2(3m^2\alpha+4i\omega))h_{rx}(r)\\
    &-8r^3(-2r^3+2r_0^3+m^2r^2\alpha-2m^2r\beta)(q^2+6r^2+2m^2\beta-2r(m^2\alpha+i\omega))h_{vr}(r)\\
    &+8r^3(q^2r+2r^3-2r_0^3+2m^2r\beta-r^2(m^2\alpha+2i\omega))h_{vv}(r)\\
    &-4iqr^2(6r_0^3+r(m^2(r\alpha-4\beta)+4ir\omega))h_{vx}(r)\\
    &+2(8r^6-4r_0^6+8m^2rr_0^3\beta-4r^5(m^2\alpha+2i\omega)-2r^2(m^2r_0^3\alpha+2m^4\beta^2-ir_0^3\omega)\\
    &+r^3(-4r_0^3+2m^4\alpha\beta-4im^2\beta\omega)+r^4(4\omega^2+m^2(4\beta+3i\alpha\omega)))h_{xx}(r)\\
    &+2(8r^6-4r_0^6+8m^2rr_0^3\beta+2q^2r(2r^3-2r_0^3-m^2r^2\alpha+2m^2r\beta)\\
    \end{aligned}
\end{equation}
\begin{equation}
\begin{aligned} 
    &-4r^5(m^2\alpha+2i\omega)-2r^2(m^2r_0^3\alpha+2m^4\beta^2-ir_0^3\omega)\\
    &+r^3(-4r_0^3+2m^4\alpha\beta-4im^2\beta\omega)
    +r^4(4\omega^2+m^2(4\beta+3i\alpha\omega)))h_{yy}(r)\\
    &+2r^2(2r^3-2r_0^3-m^2r^2\alpha+2m^2r\beta)^3h_{rr}'(r)\\
    &+4iqr^2(2r^3-2r_0^3-m^2r^2\alpha+2m^2r\beta)^2h_{rx}'(r)\\
    &+8r^3(2r^3-2r_0^3-m^2r^2\alpha+2m^2r\beta)^2h_{vr}'(r)\\
    &+4iqr^2(2r^3-2r_0^3-m^2r^2\alpha+2m^2r\beta)^2h_{rx}'(r)\\
    &+8r^3(2r^3-2r_0^3-m^2r^2\alpha+2m^2r\beta)^2h_{vr}'(r)\\
    &+8r^4(2r^3-2r_0^3-m^2r^2\alpha+2m^2r\beta)h_{vv}'(r)\\
    &+8iqr^3(2r^3-2r_0^3-m^2r^2\alpha+2m^2r\beta)h_{vx}'(r)\\
    &-r(2r^3-2r_0^3-m^2r^2\alpha+2m^2r\beta)(6r_0^3+r(m^2(r\alpha-4\beta)-8ir\omega))(h_{xx}'(r)+h_{yy}'(r))\\
    &-2r^2(2r^3-2r_0^3-m^2r^2\alpha+2m^2r\beta)^2(h_{xx}''(r)+h_{yy}''(r))=0.
\end{aligned}
\end{equation}
With the near-horizon expansion of $h_{\mu\nu}(r)=(r-r_h)^{\lambda}(\sum_{i=0}^{\infty}(r-r_h)^i h_{\mu\nu}^{(i)}(r_h))$, we find the $vv$-component equation of motion to be at leading order 
\begin{equation}
    2 h_{vv}^{(0)} r_h (q^2-2ir_h\omega)+2ih_{vx}^{(0)} q+i(h_{xx}^{(0)} + h_{yy}^{(0)})\omega(-3{r_h}^2+m^2r_h\alpha-m^2\beta-2ir_h\omega).
\end{equation}
This constraint equation vanishes at
\begin{equation}
\begin{aligned}
    \omega &= \frac{i(3{r_h}^2-m^2 r_h\alpha+m^2\beta)}{2r_h},\quad q^2=-3{r_h}^2 + m^2 r_h\alpha - m^2\beta, \\
    \omega&=q=0,
\end{aligned}
\end{equation}
and therefore these points correspond to the leading and sub-leading skipped poles respectively. One can further check that the leading order frequency satisfies the Matsubara frequency on the upper-half plane, i.e. $\omega = 2i\pi T$ by inserting the background temperature $T$. In the limit $m\rightarrow 0$, the locations of these points are consistent with the results of the massless theory in 3+1 dimensions (i.e. \eqref{sounps}).

\paragraph{Diffusive mode}

Similarly, the diffusive mode equations of motion derived from the gravitational perturbation are
\begin{equation}
\begin{aligned}
    0 &= -2ir(2r^3-2r_0^3+2m^2r\beta-r^2(m^2\alpha+i\omega))\omega h_{ry}(r)
       + 2(q^2r+2r^3-2r_0^3-2ir^2\omega) h_{vy}(r)\\
      &+r(2q\omega h_{xy}(r)+r(-i(2r^3-2r_0^3-m^2r^2\alpha+2m^2r\beta)\omega h_{ry}'(r)+2ir\omega h_{vy}'(r)\\
      &+(-2r^3+2r_0^3+m^2r^2\alpha-2m^2r\beta)h_{vy}''(r))),\\
    0 &= r(q^2+m^2(r\alpha-2\beta)+2ir\omega)h_{ry}(r)-2rh_{vy}(r)-2iqh_{xy}(r)
      +ir^3\omega h_{ry}'(r)\\
      &+iqrh_{xy}'(r)+r^3h_{vy}''(r),\\
    0 &= qr(4ir^3+2ir_0^3+r^2(2\omega-im^2\alpha))h_{ry}(r) 
       + (4r^3+8r_0^3-4m^2r\beta+r^2(m^2\alpha-4i\omega))h_{xy}(r)\\
      &+r(-6r_0^3+r(m^2(4\beta-r\alpha)+4ir\omega))h_{xy}'(r)+r^2(iq(2r^3-2r_0^3-m^2r^2\alpha+2m^2r\beta)h_{ry}'(r)\\
      &+2iqrh_{vy}'(r)+(-2r^3+2r_0^3+m^2r^2\alpha-2m^2r\beta)h_{xy}''(r)).
\end{aligned}
\end{equation}
With the near-horizon expansion $h_{\mu\nu}(r)=(r-r_h)^{\lambda}(\sum_{i=0}^{\infty}(r-r_h)^i h_{\mu\nu}^{(i)}(r_h))$, we find by eliminating the $h_{ry}$ terms the leading and sub-leading order skipped poles at
\begin{equation}
\begin{aligned}
    \omega&=0,\quad q^2=-m^2(r\alpha-2\beta),\\
    \omega&=\frac{i(-3r_h^2+m^2r_h\alpha-m^2\beta)}{2r_h},\quad
    q^2=\frac{-6r_h^3\alpha+12r_h^2\beta-2m^2r_h\alpha\beta+4m^2\beta^2 \pm\sqrt{2Q}}{2r_h\alpha},
\end{aligned}
\end{equation}
where
\begin{equation}\notag
\begin{aligned}
    Q&=72r_h^6\alpha^2-32m^4r_h\alpha\beta^3+8m^4\beta^4+4m^2r_h^2\beta^2(11m^2\alpha^2+12\beta)
    \\
    &-m^2r_h^3\alpha\beta(25m^2\alpha^2+144\beta)-3r_h^5(13m^2\alpha^3+48\alpha\beta)+r_h^4(5m^4\alpha^4+132m^2\alpha^2\beta+72\beta^2).
\end{aligned}
\end{equation}
The sub-leading order frequency satisfies the Matsubara frequency on the lower-half plane, i.e. $\omega = -2i\pi T$. In the $m\rightarrow 0$ limit, these points reduce to
\begin{equation}
\begin{aligned}
    \omega &= 0, \quad q = 0,\\
    \omega &= -\frac{3ir_0}{2}, \quad \{q_1^2 = 3r_0^2, \;\; q_2^2 = \frac{3r_0(-3r_0\alpha+4\beta)}{\alpha}\}.
\end{aligned}
\end{equation}
The leading order and the first pair of the sub-leading order (i.e. $q_1$) points are consistent with the results of the massless theory in 3+1 dimensions (i.e. \eqref{diffps}). The second pair of the sub-leading order points (i.e. $q_2$) emerges from the extra degree of freedom brought by the mass of the graviton. Such kind of extra skipped poles were also detected in topological massive gravity in \cite{Liu:2020yaf}.

\section{Conclusion and discussion}\label{sec4}

We have seen from previous examples and our gauged models with or without mass that in general, the pole-skipping takes place at Matsubara frequencies $\omega_*=-2\pi i (n-l)T, (n=1,2,3,...)$ for the holographic correlators of spin-$l$ fields, which is consistent with the spin-frequency relation observed in \cite{Wang:2022xoc}. 
For the corresponding pole-skipping wave numbers $q=q_{n}^*$, we discover with massive form models and the massive gravity models that massive correlators have doubled pole-skipping points in certain channels compared with their massless counterparts at sub-leading orders ($n\ge 2$).
The main results of the pole-skipping points of form models and massive gravity models are summarized in table \ref{pformtable} and \ref{gravitytable}.
\renewcommand\arraystretch{2} 
\begin{table}[h]
\caption{The pole-skipping values of squared wave number $q_*^2$ for $p$-form fields.}\label{pformtable}
\resizebox{\linewidth}{!}{
\begin{tabular}{|ll|l|l|}
\hline
\multicolumn{2}{|l|}{$p$-form pole-skipping point }          & First order ($\omega=0$) & Second order ($\omega=-i 2\pi T$)                                                                                                                  \\ \hline
\multicolumn{1}{|l|}{\multirow{2}{*}{Pure gauge mode}}      & massless & None                       & None                                                                                                                                                    \\ \cline{2-4} 
\multicolumn{1}{|l|}{}                                   & massive  & $-m^2 h(r_0)$              & $(d-2p+2)\pi T h'(r_0)+h(r_0)\left(-m^2+\frac{2\pi T Z'(r_0)}{Z(r_0)}\right)$                                                                           \\ \hline
\multicolumn{1}{|l|}{\multirow{2}{*}{Longitudinal mode}} & massless & $0$                        & $\pi T \left( (d-2p)h'(r_0)+\frac{2h(r_0) Z'(r_0)}{Z(r_0)}\right)$                                                                                      \\ \cline{2-4} 
\multicolumn{1}{|l|}{}                                   & massive  & $-m^2 h(r_0)$              & $- m^2 h(r_0)- \pi T h'(r_0) \pm \pi T\sqrt{\left[(d-2p+1)h'(r_0)+2h(r_0)\frac{Z'(r_0)}{Z(r_0)}\right]^2+\frac{4m^2}{\pi T}h(r_0)h'(r_0)}$ \\ \hline
\multicolumn{1}{|l|}{\multirow{2}{*}{Transverse mode}}   & massless & None                       & $-\pi T \left( (d-2p)h'(r_0)+\frac{2h(r_0) Z'(r_0)}{Z(r_0)}\right)$                                                                                     \\ \cline{2-4} 
\multicolumn{1}{|l|}{}                                   & massive  & None                       & $-\pi T \left( (d-2p)h'(r_0)+\frac{2h(r_0) Z'(r_0)}{Z(r_0)}\right)-m^2 h(r_0)$                                                                          \\ \hline
\end{tabular}}
\end{table}
\begin{table}[h]
\caption{The pole-skipping values of squared wave number $q_*^2$ for gravitational fluctuations.}\label{gravitytable}
\resizebox{\textwidth}{!}{%
\begin{tabular}{|ll|l|l|l|}
\hline
\multicolumn{2}{|l|}{Gravity pole-skipping point ($d=2$)}                        & First order ($\omega=i2\pi T$)        & Second order ($\omega=0$)                   & Third order ($\omega=-i2\pi T$)                                                                                                                                                                                                                                                                                 \\ \hline
\multicolumn{1}{|l|}{\multirow{3}{*}{Diffusive mode}} & massless                 & None                                  & $0$                                         & $3r_0^2$                                                                                                                                                                                                                                                                                                        \\ \cline{2-5} 
\multicolumn{1}{|l|}{}                                & massive ($f=\Bar{g}$)    & None                                  & $-m^2 r_0^2$                                & $\left(-\frac{3}{2}-m^2\pm \frac{1}{2}\sqrt{81+24m^2}\right)r_0^2$                                                                                                                                                                                                                                              \\ \cline{2-5} 
\multicolumn{1}{|l|}{}                                & massive ($f\ne \Bar{g}$) & None                                  & $0$                                         & \begin{tabular}[c]{@{}l@{}}$2 q^4 r_h \alpha - 54 r_h^5 \alpha -42 m^4 r_h \alpha \beta^2 + 24 m^4 \beta^3 + r_h^4 (39 m^2 \alpha^2 + 72 \beta)+m^2 r_h^2 \beta (25 m^2 \alpha^2$ \\ $ + 96 \beta) + 4 q^2 (r_h \alpha - 2 \beta) (3 r_h^2 + m^2 \beta) -5 r_h^3 (m^4 \alpha^3 + 24 m^2 \alpha \beta) = 0$\end{tabular} \\ \hline
\multicolumn{1}{|l|}{\multirow{3}{*}{Sound mode}}     & massless                 & $-3r_0^2$                             & $0$                                         & $\pm 3ir_0^2$                                                                                                                                                                                                                                                                                                   \\ \cline{2-5} 
\multicolumn{1}{|l|}{}                                & massive ($f=\Bar{g}$)    & $-(3+m^2)r_0^2$                       & $\left(-3-m^2\pm \sqrt{9+6m^2}\right)r_0^2$ & \begin{tabular}[c]{@{}l@{}}$q^8+(15+4m^2)r_0^2 q^6+3(9+7m^2+2m^4)r_0^4 q^4+(135+18m^2$\\ $-3m^4+4m^6)r_0^2 q^2+(162+135m^2-9m^4-9m^6+m^8)r_0^8=0$\end{tabular}                                                                                                                                                  \\ \cline{2-5} 
\multicolumn{1}{|l|}{}                                & massive ($f\ne \Bar{g}$) & $−3r_h^2 + m^2 r_h \alpha− m^2 \beta$ & $0$                                         & None                                                                                                                                                                                                                                                                                                            \\ \hline
\end{tabular}%
}
\end{table}
\renewcommand\arraystretch{1}

\begin{figure}[htbp]
    \centering
    \subfigure[]{
    \begin{minipage}[b]{.4\linewidth}
    \centering
    \includegraphics[scale=0.4]{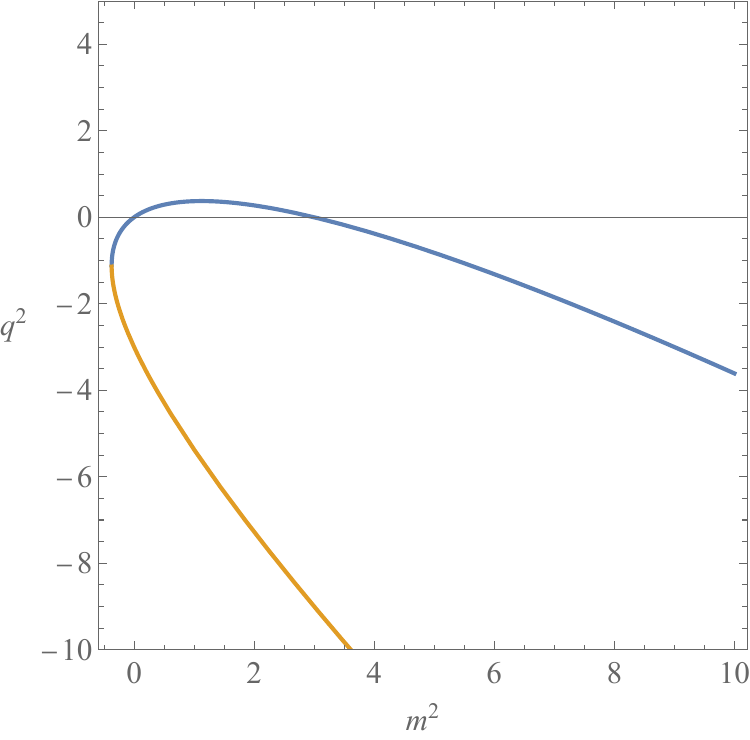}
    \end{minipage}
    }
    \subfigure[]{
    \begin{minipage}[b]{.4\linewidth}
    \centering
    \includegraphics[scale=0.4]{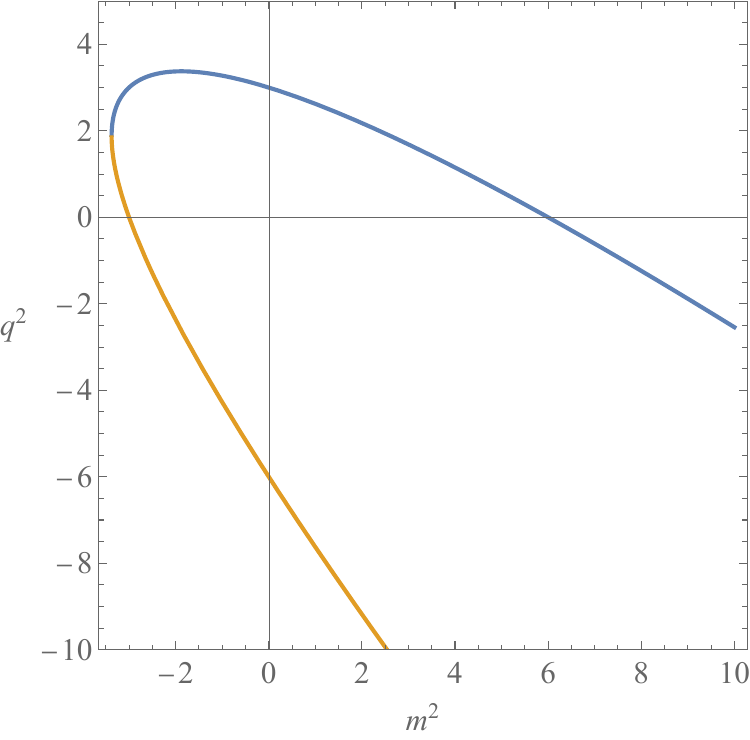}
    \end{minipage}
    }
    
    \subfigure[]{
    \begin{minipage}[b]{.4\linewidth}
    \centering
    \includegraphics[scale=0.4]{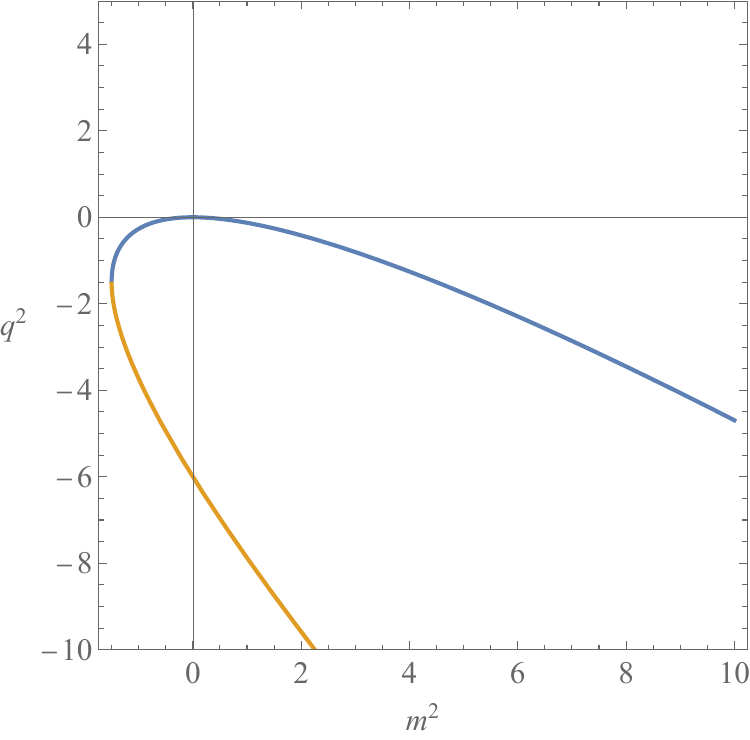}
    \end{minipage}
    }
    \subfigure[]{
    \begin{minipage}[b]{.4\linewidth}
    \centering
    \includegraphics[scale=0.4]{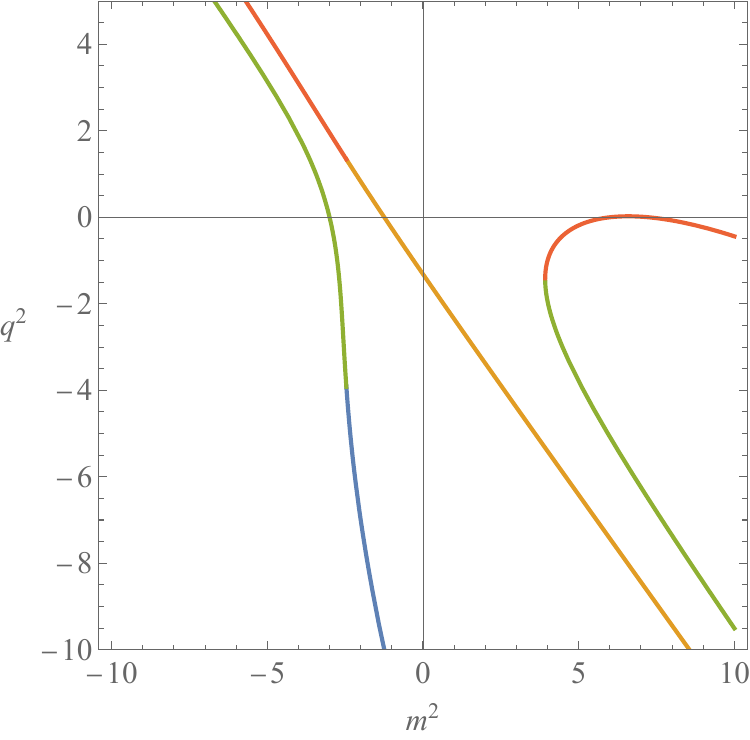}
    \end{minipage}
    }
    \caption{The pole-skipping values of squared wave number $q^2$ as functions of squared mass $m^2$ for (a) the longitudinal mode of $1$-form fields in AdS$_4$ (at order $\omega=-2i\pi T$), (b) the diffusive mode of metric fluctuations (at order $\omega=-2i\pi T$ in massive gravity model of $f=\Bar{g}$), (c) the sound mode of metric fluctuations (at order $\omega=0$ in massive gravity model of $f=\Bar{g}$), (d) the sound mode of metric fluctuations (at order $\omega=-2i\pi T$ in massive gravity model of $f=\Bar{g}$). $r_0$ is set to be $1$.}
    \label{qrealornot}
\end{figure}

The pole-skipping results for $q_*^2$ are in general complex, so the skipped poles we found are not all physical since a physical wave number must be real, although a physical frequency can have a non-zero imaginary part which describes the decay of the mode. 
It is interesting to note that as the mass $m$ varies, the wave number may move from the non-physical region (complex $q_*$) to the physical region (real $q_*$) and vise versa.
For instance, in our massive gravity model of $f=\Bar{g}$, the sound mode have four pole-skipping points at the third order $\omega=-2i\pi T$.
In the mass range $m^2\in (-\infty,-3)$, two of the special points have positive $q_*^2$ (real $q_*$), while for $m^2\in \left(-3,3-3\sqrt{2}\right)\bigcup \left(6,3+3\sqrt{2}\right)$ only one $q_*^2$ is positive, and for other values of $m^2$ none of these are positive.
It is interesting to explore more on the physical implication of these mass ranges in future works.
Figure \ref{qrealornot} shows four examples for the dependence of $q_*^2$ on $m^2$.

According to the Stueckelberg formalism, the massless limit of a massive theory is equivalent to the massless theory plus a decoupled auxiliary field. 
Therefore, as we have observed earlier, in zero mass limit, the first half of the pole-skipping points reduce to the massless results while the other half reduce to the pole-skipping points of the auxiliary field introduced by Stueckelberg formalism. 
Because more degrees of freedom are associated to a massive mode than the its massless counterpart, it is reasonable to expect that this doubling phenomenon of pole-skipping is in fact universal in the holographic correlators of models with gauge symmetry breaking (by a non-zero mass).
To conclude, our results are in the affirmative for the universal relationship between the double pole-skipping phenomenon and gauge symmetry/symmetry breaking, valid not only for form models and massive gravity models but also possibly for other gauge theories in the bulk.

\section*{Acknowledgement}

We thank Teng-Zhou Lai for collaboration at an early stage of this work and his helpful suggestions. 
We also thank Hyun-Sik Jeong, Pan Li, Yan Liu, Yi Ling and Xuan-Ting Ji for their helpful discussion and advice. 
This work was supported by Project 12347183, 12035016 and 12275275 supported by the National Natural Science Foundation of China.
It is also supported by Beijing Natural Science Foundation under Grant No. 1222031.

\appendix

\section{An alternative method for longitudinal modes}\label{appendixA}

In this appendix, we establish a new method to compute the pole-skipping points of longitudinal modes.
This method not only gives the same results in the massless case as the conventional one first proposed in \cite{Blake:2019otz}, but also applies to the massive case.
We demonstrate here the method for the $1$-form field as an example, the generalization to arbitrary integer $p$ is straightforward.

\subsection{Massless longitudinal modes}\label{appendixA1}
The equations of motion for the massless longitudinal mode are three first-order ODEs coupled by gauge invariants $F_{vr}$, $F_{vx}$ and $F_{rx}$ as follows,
\begin{align}
    \partial_r(U(r) g^{rv}g^{vr}F_{vr})-iqU(r)g^{xx}g^{vr}F_{rx}=0\\
    -iqU(r)g^{xx}g^{rr}F_{rx}+i\omega U(r)g^{vr}g^{rv}F_{vr}-iqU(r)g^{xx}g^{rv}F_{vx}=0\label{eom2}\\
    \partial_r(U(r) g^{rv}g^{xx}F_{vx})+\partial_r(U(r) g^{rr}g^{xx}F_{rx})-i\omega U(r)g^{vr}g^{xx}F_{rx}=0
\end{align}
where we have defined $U(r)=\sqrt{-g}Z(r)$ and $g^{rv}=g^{vr}=1$, $g^{xx}=\frac{1}{h(r)}$, $g^{rr}=r^2 f(r)$.
These three gauge invariants also satisfy a single constraint
\begin{align}
    -i\omega F_{rx}-\partial_{r} F_{vx}+iq F_{vr}=0.
\end{align}
To simplify these equations, we note that the equation \eqref{eom2} does not contain any derivative, such that we could reduce one of the variables by solving $F_{rx}$ (or $F_{vr}$, $F_{vx}$) from \eqref{eom2}:
\begin{align}
    F_{rx}=\frac{-q F_{vx}+\omega h(r) F_{vr}}{q r^2 f(r)},
\end{align}
and substituting into the other two equations of motion and the constraint.
One of the three remaining equations is redundant and we arrive at two independent first-order ODEs:
\begin{align}\label{master}
   \psi'(r)=A(r)\psi(r)
\end{align}
where we denote $\psi=(F_{vr},F_{vx})^T$ and $A(r)=\begin{pmatrix}
  \frac{i \omega}{r^{2} f(r)}-\frac{d h^{\prime}(r)}{2 h(r)}-\frac{z^{\prime}(r)}{z(r)}& -\frac{i q}{r^{2} f(r) h(r)}\\
   i q-\frac{i \omega^{2} h(r)}{q r^{2} f(r)}& \frac{i \omega}{r^{2} f(r)}
\end{pmatrix}$.

$A(r)$ has a first-order pole at $r=r_0$ and can be series expanded as 
\begin{align}
   A(r)=\frac{A_0}{r-r_0}+A_1+A_2 (r-r_0)+ O(r-r_0)^2,
\end{align}
where $A_0$, $A_1$, $A_2$ are matrix valued constant of $r$.
We also assume the power series solution for $\psi$
\begin{align}
   \psi(r)=(r-r_0)^\lambda(\psi_0+\psi_1 (r-r_0)+ O(r-r_0)^2).
\end{align}
with $\psi_0$ and $\psi_1$ are vector-valued constant of $r$.
Note that $\psi_0$ is assumed not equal to zero, in order for the leading exponent to be $\lambda$.
Keep to the first two orders, \eqref{master} reads
\begin{align}
   (\lambda I -A_0)\psi_0 (r-r_0)^{\lambda-1}+(((\lambda+1) I -A_0)\psi_1-A_1 \psi_0) (r-r_0)^{\lambda}+O(r-r_0)^{\lambda+1}=0.
\end{align}
Then we could solve \eqref{master} order by order.
The leading order equation
\begin{align}\label{first}
   (\lambda I -A_0)\psi_0=0
\end{align}
tells us that $\lambda$ must equal to one of the eigenvalues of $A_0$, otherwise $\psi_0=0$ which is in contrary to our assumption.
The sub-leading order equation reads
\begin{align}\label{second}
   ((\lambda+1) I -A_0)\psi_1-A_1 \psi_0=0.
\end{align}
Let $\lambda_1$ and $\lambda_2$ be the two eigenvalues of $A_0$. 
If we choose one of the eigenvalues, say $\lambda=\lambda_1$, the matrix $(\lambda_1+1) I -A_0$ is invertible unless $\lambda_2-\lambda_1=1$.
Therefore, in general, one can not represent $\psi_1$ in terms of $\psi_0$ from \eqref{second} when the difference between the two eigenvalues happens to be $1$ (if $\lambda$ is chosen to be the smaller eigenvalue).

However, we will show that even if $\lambda_2-\lambda_1=1$, there is still one possibility that one can solve $\psi_1$ from \eqref{second}.
Let $u_1$ and $u_2$ be the two eigenvectors that correspond to $\lambda_1$ and $\lambda_2=\lambda_1+1$ respectively,
\begin{align}\label{eigen}
   \left\{\begin{matrix}
 A_0 u_1=\lambda_1 u_1\\
 A_0 u_2=(\lambda_1+1) u_2
\end{matrix}\right.
\end{align}
$u_1$ and $u_2$ are linearly independent. 
Thus, any 2-d vector like $\psi_0$ and $\psi_1$ can be expressed as a linear combination of $u_1$ and $u_2$:
\begin{align}\label{coef}
   \left\{\begin{matrix}
 \psi_0=c^1_0 u_1+c^2_0 u_2\\
 \psi_1=c^1_1 u_1+c^2_1 u_2
\end{matrix}\right.
\end{align}
We first derive the condition that $c^1_0$ and $c^2_0$ need to satisfy from the leading equation \eqref{first}.
Set $\lambda=\lambda_1$ and put \eqref{eigen} and \eqref{coef} into \eqref{first}, we obtain 
\begin{align}
   (\lambda I -A_0)\psi_0=(\lambda_1 I -A_0)(c^1_0 u_1+c^2_0 u_2)=-c^2_0 u_2=0,
\end{align}
which demands $c^2_0=0$ and $\psi_0=c^1_0 u_1$ with $c^1_0$ be a free parameter in the solution.
Then we try to solve $c^1_1$ and $c^2_1$ from the sub-leading equation \eqref{second}.
Put \eqref{eigen} and \eqref{coef} into \eqref{second}, we find
\begin{align}
  0&= ((\lambda+1) I -A_0)\psi_1-A_1 \psi_0\\
   &=((\lambda_1+1) I -A_0)(c^1_1 u_1+c^2_1 u_2)-A_1 (c^1_0 u_1+c^2_0 u_2)\\
   &=c^1_1 u_1-c^1_0 A_1 u_1.
\end{align}
This equation requires the vector $u_1$ and $A_1 u_1$ to be linear dependent or else $c^1_1=c^1_0=0$ (which contradicts to the assumption $\psi_0 \ne 0$).
Therefore the condition that one can solve $\psi_1$ ($c^1_1$) is that 
\begin{align}\label{k2ps}
   A_1 u_1= \Tilde{\lambda} u_1,
\end{align}
which means $u_1$ is a common eigenvector of $A_0$ and $A_1$.
We propose the condition \eqref{k2ps} together with $\lambda_2=\lambda_1+1$ are the conditions that determine the first order pole-skipping point for the diffusive channel current-current retarded correlator and we show below that they are consistent with the results in the literature. 
It is important to note that when the two conditions are fulfilled, there are two free parameters in the series solution: $c_0^1$ and $c_1^2$, which correspond to two independent solutions whose linear combinations constitute the non-unique ingoing solution.

In the current massless case,
\begin{align}
    A_0=\begin{pmatrix}
  \frac{i \omega}{r_0^{2} f'(r_0)}& -\frac{i q}{r_0^{2} f'(r_0) h(r_0)}\\
   -\frac{i \omega^{2} h(r_0)}{q r_0^{2} f'(r_0)}& \frac{i \omega}{r_0^{2} f'(r_0)}
   \end{pmatrix}
\end{align}
and
\begin{align}
    A_1=\begin{pmatrix}
  -\frac{d h^{\prime}(r_0)}{2 h(r_0)}-\frac{Z^{\prime}(r_0)}{Z(r_0)}-\frac{i \omega\left(4 f^{\prime}(r_0)+r_0 f^{\prime \prime}(r_0)\right)}{2 r_0^{3} f^{\prime}(r_0)^{2}} & \frac{i q\left(2 r_0 f^{\prime}(r_0) h^{\prime}(r_0)+h(r_0)\left(4 f^{\prime}(r_0)+r_0 f^{\prime \prime}(r_0)\right)\right)}{2 r_0^{3} h(r_0)^{2} f^{\prime}(r_0)^{2}} \\
i q+\frac{i \omega^{2}\left(-2 r_0 f^{\prime}(r_0) h^{\prime}(r_0)+h(r_0)\left(4 f^{\prime}(r_0)+r_0 f^{\prime \prime}(r_0)\right)\right)}{2 q r_0^{3} f^{\prime}(r_0)^{2}} & -\frac{i \omega\left(4 f^{\prime}(r_0)+r_0 f^{\prime \prime}(r_0)\right)}{2 r_0^{3} f^{\prime}(r_0)^{2}}
   \end{pmatrix}
\end{align}
The eigenvalues of $A_0$ are $\lambda_1=0$ and $\lambda_2=\frac{2i\omega}{r_0^2 f'(r_0)}$.
We choose the leading exponent $\lambda=\lambda_1=0$ because this corresponds to the outgoing solution near the horizon.
In the special case $\omega=\frac{r_0^2 f'(r_0)}{2i}=-i 2\pi T$, we have $\lambda_2-\lambda_1=1$, satisfying the above description.
To obtain the special value for $a$, we use the condition \eqref{k2ps}.
We find 
\begin{align}
   u_1&= (\frac{iq}{2\pi T h(r_0)},1)^T\\
   A_1 u_1&= (-\frac{i q\left((d-1) Z(r_0) h^{\prime}(r_0)+2 h(r_0) Z^{\prime}(r_0)\right)}{4 \pi T h(r_0)^{2} Z(r_0)},-\frac{q^{2}+\pi T h^{\prime}(r_0)}{2 \pi T h(r_0)})^T
\end{align}
The proportionality condition of the two vectors demands that 
\begin{align}\label{mvps}
   q^2=\pi T \left( (d-2)h'(r_0)+\frac{2h(r_0) Z'(r_0)}{Z(r_0)}\right)
\end{align}

Finally, we conclude that the first-order pole-skipping point is at $(\omega=-i 2\pi T, q^2=\pi T \left( (d-2)h'(r_0)+\frac{2h(r_0) Z'(r_0)}{Z(r_0)}\right))$, which is consistent with previous literature \cite{Blake:2019otz,Natsuume:2019sfp,Natsuume:2019xcy}.
\vspace{.5em}

In addition, by applying the following iteration relationship, 
\begin{align}
\sum_{k=0}^{n}A_{n-k}\psi_{k}-(\lambda+n)I\psi_{n}=0
\end{align}
we can analyze the higher-order pole-skipping points in this case. For the second-order, which means $\omega=-i 4\pi T$, the coefficient matrix in front of $\psi_{1}$ is invertible. So $\psi_{1}$ can be directly expressed by $\psi_{0}$, 
\begin{align}
\psi_{1}=-(A_{0}-I)^{-1}A_{1}\psi_{0}
\end{align}
 as the first order we noticed that the eigenvector $u_{1}$ of $A_{0}$ is also the one of the eigenvector of the matrix $A_{2}-A_{1}(A_{0}-I)^{-1}A_{1}$. so we can take the same way to calculate the second order pole-skipping point. Or equivalently applying the matrix $P=\left\{u^{T}_{1},u^{T}_{2}\right\}$ to do similar transformation for the above matrix: $B=P^{-1}\left[A_{2}-A_{1}(A_{0}-I)^{-1}A_{1}\right]P$, then requiring the element $B_{21}$ vanish, which will give us the following second-order pole-skipping points 
 \begin{align}
  q^{2}=&\pi T\left[(d-4)h'(r_{0})+\frac{2h(r_0)Z'(r_0)}{Z(r_0)}+\frac{8h(r_0)}{r_{0}}\right]+\frac{r_{0}^{2}}{2}h(r_0)f''(r_0)\pm \sqrt{\gamma}\\
 \gamma=&\left[d\pi T h'(r_0)+\frac{h(r_0)}{2r_0 Z(r_0)}\left[Z(r_0)(16\pi T+r^{3}_{0}f''(r_0))-4\pi r_0 T Z'(r_0)\right]\right]^{2}\notag\\
 &-16\pi^2 T^2\left(\frac{h(r_0)h'(r_0)Z'(r_0)}{Z(r_0)}+\frac{d-2}{2}h(r_0)h''(r_0)-h^{2}(r_0)\frac{(Z'(r_0))^2-Z''(r_0)Z(r_0)}{Z^{2}(r_0)}\right)
 \end{align}
 now we wish to check this result explicitly in a particular example. Considering a planar AdS$_{d+2}$-Schwarzschild spacetime, then the metric is given by $f(r)=1-(\frac{r_0}{r})^{d+1}, h(r)=r^2$ and for simplicity taking $Z(\Phi)=1$. For this case,the second pole-skipping points locate at the following place
 \begin{align}
 q^2=\left(-1\pm\sqrt{d^2-3d+3}\right)(d+1)r^{2}_{0}
 \end{align}
which is also consistent with the result from the literature\cite{Blake:2019otz}.

Actually we can find three or higher-order pole-skipping points along the same approach, due to the calcualtion is so complicated, here we just give a sketch. When $\omega=-i2\pi T n$, then we can 
 easily represent $\psi_{1}, \cdots, \psi_{n-1}$ in a specific form of $\psi_{0}$, as the matrix in front of $\psi_{m}(m=1,\cdots,n-1)$ is invertible. so the vector $u_{1}$ is again the eigenvector of the matrix $B_{n}$，which is defined as
 \begin{align}
 B_{n}&\coloneqq A_{n}+A_{n-1}\tilde{B}_{1}+\cdots+A_{1}\tilde{B}_{n-1}\notag\\ \tilde{B}_{1}&=-(A_{0}-I)^{-1}A_{1}\notag\\
 \tilde{B}_{2}&=-(A_{0}-2I)^{-1}\left(A_{2}-A_{1}(A_{0}-I)^{-1}A_{1}\right)\notag\\
 \tilde{B}_{3}&=-(A_{0}-3I)^{-1}\left[A_{3}-A_{2}(A_{0}-I)^{-1}A_1 -A_{1}(A_0 -2I)^{-1}\left(A_{2}-A_{1}(A_{0}-I)^{-1}A_{1}\right)\right]\notag\\
 \cdots
 \end{align}
 then requiring the element $\left(P^{-1}B_{n}P\right)_{21}$ vanish , we can get $n$ different pole-skipping points.

\subsection{Massive longitudinal modes}\label{appendixA2}
In the massive case, we derive the equation of motion for the gauge invariants of the Stueckelberg formalism and then calculate the corresponding first-order pole-skipping point by using a similar method as in the massless case.

In this new method, we focus on the equations for gauge invariant variables instead of the components of the form field as in the main text.
By extremizing the action \eqref{stuck}, we find the equation of motion and the constraint for $F$ are\footnote{$J$ is the new gauge invariant after the auxiliary scalar $\phi$ is introduced: $J_\mu=A_\mu +\frac{1}{m}\partial_\mu \phi$}
\begin{equation}\label{eomF}
    \begin{aligned}
    \nabla_\mu (Z(r) F^{\mu \nu})&=m^2 Z(r) J^\nu,\\
    \partial_{[\rho} F_{\mu \nu]}&=0,
    \end{aligned}
\end{equation}
while the equation of motion and the constraint for $J$ are respectively
\begin{equation}\label{eomJ}
    \begin{aligned}
    \nabla_\mu (Z(r) J^{\mu })&=0,\\
    \partial_{[\mu} J_{\nu]}&=F_{\mu \nu}.
    \end{aligned}
\end{equation}

In the following calculation we will work with a general boundary spatial $d$ dimension. The diffusive channel $(F_{vr}, F_{vx}, F_{rx}, J_v, J_r, J_x)$ and the transverse channel $(F_{vj}, F_{rj}, F_{xj}, J_j)$ ($j$ denotes transverse coordinates) decouple again in the equations.
There are $8$ equations for the diffusive channel in \eqref{eomF} and \eqref{eomJ} and we have $6$ variables.
$2$ of the equations do not contain $r$ derivative
\begin{align}
   0&=\partial_\mu (U(r) g^{\mu \rho} g^{r \sigma} F_{\rho \sigma})-m^2 U(r) g^{r \rho} J_{\rho}\notag\\&=U(r)\left[i\omega F_{v r}-i q \frac{r^2 f(r)}{h(r)}F_{r x}-\frac{i q}{h(r)}F_{v x}-m^{2}(r^{2}f(r) J_{r}+J_{v})\right],\\
   0&=F_{vx}-\partial_{[v} J_{x]}= F_{vx} +i q J_v +i \omega J_x
\end{align}
so that can be used to cancel $2$ variables and the number of equations can be reduced by $2$, here we take the same definition as the previous massless case: $U(r)\equiv\sqrt{-g}Z(r)$. We choose to cancel $F_{rx}$ and $J_x$.
Then there are $6$ equations for $4$ variables $(F_{vr}, F_{vx}, J_v, J_r)$.
We find $2$ of these equations are redundant.
Thus $4$ independent equations are left for $4$ variables, so eventually \eqref{eomF} and \eqref{eomJ} can be simplified into the form of \eqref{master}: 
\begin{align}
   \psi'(r)=A(r)\psi(r)
\end{align}
 in which $A(r)$ is a $4 \times 4$ matrix and $\psi=(F_{vr}, F_{vx}, J_v, J_r)^T$,and the square matrix $A(r)$ is given by
 \begin{small}
    \begin{align}
A(r)=\frac{1}{r^2 f(r)}\begin{pmatrix}
i\omega-r^{2}f(r)\frac{\partial_{r}U(r)}{U(r)}& \frac{-i q}{h(r)} & -m^2 & 0 \\
\frac{-i \omega^{2} h(r)}{q}+i q\,r^{2}f(r) & i\omega &\frac{m^2\omega h(r)}{q} & \frac{m^2\omega h(r)}{q} r^2 f(r)\\
-r^2 f(r) & 0 & 0 & -i\omega\,r^2 f(r)\\
1 & \frac{q}{\omega h(r)} &\frac{i q^2}{\omega h(r)}-\frac{\partial_{r}U(r)}{U(r)}& 2i\omega-2rf(r)-r^{2}f'(r)-r^{2}f(r)\frac{\partial_{r}U(r)}{U(r)}
\end{pmatrix}
 \end{align} 
 \end{small}
 , in which $\frac{\partial_{r}U(r)}{U(r)}=\left(\frac{d h'(r)}{2h(r)}+\frac{Z'(r)}{Z(r)}\right)$.

In the first order, we have
\begin{align}
    A_0=\frac{1}{r_0^2 f'(r_0)}\begin{pmatrix}
  i \omega& -\frac{i q}{h(r_0)}&-m^2&0\\
   -\frac{i \omega^{2} h(r_0)}{q}& i \omega & \frac{m^2 \omega h(r_0)}{q}&0\\
   0&0&0&0\\
   1&\frac{q}{\omega h(r_0)}& \frac{i q^2}{\omega h(r_0)}-\frac{dh'(r_0)}{2h(r_0)}-\frac{Z'(r_0)}{Z(r_0)}& 2i\omega-r_0^2 f'(r_0)
   \end{pmatrix}
\end{align}
and its eigenvalues are 
\begin{align}
   \lambda_1=\lambda_2=0, \quad \lambda_3=\frac{i\omega}{2 \pi T}-1, \quad \lambda_4=\frac{i\omega}{2 \pi T}.
\end{align}
As deduced in the massless case, the leading exponent $\lambda$ of $\psi$ must be equal to one of the eigenvalues. 
And we know the ingoing solution for $\psi$ corresponds to $\lambda=\lambda_1=\lambda_2=0$, so we choose $\lambda=0$ and try to solve this ingoing solution order by order with the ansatz
\begin{align}
   \psi(r)=\psi_0+\psi_1 (r-r_0)+ O(r-r_0)^2.
\end{align}
As the same in the massless case, the leading and sub-leading order equations are
\begin{align}\label{eqmass}
   &(\lambda I -A_0)\psi_0=0,
   &((\lambda+1) I -A_0)\psi_1-A_1 \psi_0=0.
\end{align}
The first-order pole-skipping for $\omega$ is determined by the sub-leading equation to be $\omega=-i 2\pi T$ because at this value there exists an eigenvalue whose difference with $\lambda$ is $1$ such that the matrix $(\lambda+1) I -A_0$ is not invertible.

To solve $\psi_1$ when $\omega=-i 2\pi T$, we need a more careful calculation.
$A_0$ has only three linearly independent eigenvectors at $\omega=-i 2\pi T$:
\begin{scriptsize}
\begin{align}
   u_1&= (0,0,0,1)^T\\
   u_3&= (\frac{1}{4}\left( \frac{q^2+m^2 h(r_0)+d \pi T h'(r_0)}{\pi T h(r_0)}+\frac{2 Z'(r_0)}{Z(r_0)}\right), -\frac{i\left((q^2-m^2 h(r_0)+d \pi T h'(r_0))Z(r_0)+2\pi T h(r_0) Z'(r_0)\right)}{2 q Z(r_0)}, 1, 0)^T\\
   u4&=(-\frac{iq}{2\pi T h(r_0)},1,0,0)^T,
\end{align}
\end{scriptsize}
and 
\begin{align}
   A_0 u_1 &= 0\\
   A_0 u_3 &= 0\\
   A_0 u_4 &= u_4
\end{align}
Unlike the massless case, the three eigenvectors span a linear space whose dimension is smaller than the dimension of matrix $A_0$. 
This means $A_0$ can not be fully diagonalized.
In order to express $\psi_0$ and $\psi_1$ by the eigenvectors just like in the massless case, we have to find a fourth vector $u_2$ that is linear independent with all the eigenvectors.

Note that $A_0$, like all the square matrices, can be put into the Jordan standard form \cite{} by a similar transformation $P$
\begin{align}\label{Jordan}
   P^{-1}A_0 P=J
\end{align}
where $J$ is the Jordan standard form of $A_0$:
\begin{align}
   J=\begin{pmatrix}
   0& 1& 0& 0\\
   0& 0& 0& 0\\
   0& 0& 0& 0\\
   0& 0& 0& 1\\
   \end{pmatrix}
\end{align}
and $P$ is the similar transformation matrix: 
\begin{align}
   P=(u_1|u_2|u_3|u_4)
\end{align}
We use equation \eqref{Jordan} to determine $u_2$ and find
\begin{align}
   A_0 u_2=u_1.
\end{align}
The solution of $u_2$ is not unique and we do not write the solution here because the precise solution does not affect our result as long as $A_0 u_2=u_1$ holds.

After obtaining the full set of linearly independent vectors, we finally could assume
\begin{align}
   \left\{\begin{matrix}
   \psi_0=c^1_0 u_1+c^2_0 u_2+c^3_0 u_3+c^4_0 u_4\\
   \psi_1=c^1_1 u_1+c^2_1 u_2+c^3_1 u_3+c^4_1 u_4
   \end{matrix}\right.
\end{align}
where the coefficients $c^\mu_0$ and $c^\mu_1$ are what we would like to solve.
We also assume every component of $\psi_0$ is not zero in order for all the solutions of variables in $(F_{vr}, F_{vx}, J_v, J_r)^T$ to be ingoing.
Substitute this assumption into the first equation in \eqref{eqmass} and we find
\begin{equation}
\begin{aligned}
   0&=(\lambda I -A_0)\psi_0\\
   &=-A_0(c^1_0 u_1+c^2_0 u_2+c^3_0 u_3+c^4_0 u_4)\\
   &=-c^2_0 u_1-c^4_0 u_4,
\end{aligned}
\end{equation}
which demands $c^2_0=c^4_0=0$ and $c^1_0$, $c^3_0$ are free parameters.
At the same time, the second equation in \eqref{eqmass} reads
\begin{small}
\begin{equation}
\begin{aligned}
  0&= ((\lambda+1) I -A_0)\psi_1-A_1 \psi_0\\
   &=(I-A_0)(c^1_1 u_1+c^2_1 u_2+c^3_1 u_3+c^4_1 u_4)-A_1 (c^1_0 u_1+c^2_0 u_2+c^3_0 u_3+c^4_0 u_4)\\
   &=(c^1_1-c^2_1) u_1+c^2_1 u_2+c^3_1 u_3-c^1_0 A_1 u_1-c^3_0 A_1 u_3\\
   &=(c^1_1-c^2_1) u_1+c^2_1 u_2+c^3_1 u_3-c^1_0 (b^1_1 u_1+b^2_1 u_2+b^3_1 u_3+b^4_1 u_4)-c^3_0 (b^1_3 u_1+b^2_3 u_2+b^3_3 u_3+b^4_3 u_4)\\
   &=(c^1_1-c^2_1-c^1_0 b^1_1-c^3_0 b^1_3)u_1+(c^2_1-c^1_0 b^2_1-c^3_0 b^2_3) u_2+(c^3_1-c^1_0 b^3_1-c^3_0 b^3_3) u_3+(-c^1_0 b^4_1-c^3_0 b^4_3)u_4,
\end{aligned}
\end{equation}   
\end{small}
where in the fourth line we have set $A_1 u_1=b^1_1 u_1+b^2_1 u_2+b^3_1 u_3+b^4_1 u_4$ and $A_1 u_3=b^1_3 u_1+b^2_3 u_2+b^3_3 u_3+b^4_3 u_4$.
All the expressions in the brackets of the last line must vanish.
We could solve $c^\mu_1$ ($\mu=1,2,3$) from the first three brackets and $c^4_1$ is left to be free.
For the last bracket, we find $b^4_1=0$ and 
\begin{align}
 b^4_3 = \frac{i}{8 q \pi T h(r_0) Z(r_0)^{2}} &\left(Z(r_0)^{2}\left(q^{4}+m^{4} h(r_0)^{2}+2 q^{2} \pi T h^{\prime}(r_0)-(d-2)\pi^{2} T^{2} h^{\prime}(r_0)^{2} \right.\right.\notag\\
  &\left.\left.+2 m^{2} h(r_0)\left(q^{2}-\pi T h^{\prime}(r_0)\right)\right)-4(d-1)\pi^2T^2 h(r_0)Z(r_0)h^{\prime}(r_0)Z^{\prime}(r_0)\right.\notag\\ &\left.-4\pi^2T^2h(r_0)^2 Z^{\prime}(r_0)^2\right)
\end{align}
If $b^4_3\ne0$, then $c^3_0=0$ and a degree of freedom is lost in the full solution.
This means we must have $b^4_3=0$ and $c^3_0$ becomes an extra free parameter corresponding to an additional ingoing solution of the system.
Therefore we eventually find that the equation $b^4_3=0$ leads us to the first-order pole-skipping point for $q$:
\begin{align}
	q^2=-\left( m^2 h(r_0)+ \pi T h'(r_0) \pm \pi T\sqrt{\left[(d-1)h'(r_0)+2h(r_0) \frac{Z'(r_0)}{Z(r_0)}\right]^2+\frac{4m^2}{\pi T}h(r_0)h'(r_0)}\right)\label{mmp}
\end{align}
This pole-skipping point matches with the result in the section \ref{sec3.2}.

\section{The pole-skipping results for AdS-Schwarzschild black hole }\label{appendixB}

In the following four tables, we summarize the pole-skipping points of the first few orders for both the longitudinal and transverse modes in the AdS-Schwarzschild background ($f(r)=1-\left(\frac{r_0}{r}\right)^{1+d}, h(r)=r^2, Z(r)=1$).
\begin{table}[hbt!]
	\caption{The pole-skipping points for the longitudinal mode of massless $p$-form fields.}
	\label{table_ml_long}
	\centering
	\begin{tabular}{l}
		\hline
		\textbf{$1$st order} \\
		\;$\mathrm{AdS}_{d+2}$: $\omega=q^2=0$ \\
		\hline
		\textbf{$2$nd order} \\
		\;$\mathrm{AdS}_{d+2}$: $\omega=-\frac{1}{2}i(1+d)r_0,\; q^2=\frac{1}{2}(1+d)(d-2p){r_0}^2$ \\
		\hline
		\textbf{$3$rd order} \\
		\;$\mathrm{AdS}_{d+2}$: $\omega=-i(1+d)r_0,\; q^2=(1+d)\bigl(-p\pm \sqrt{d^2+p(2+p)-d(1+2p)}\bigr){r_0}^2$ \\
		\hline
		\textbf{$4$th order} \\
		\;$\omega=-\frac{3}{2}i(1+d)r_0$\\
		\quad \textbf{$1$-form}\\
		\qquad $\mathrm{AdS}_{3}$: $q^2=-{r_0}^2,-9{r_0}^2,-{r_0}^2$\\
		\qquad $\mathrm{AdS}_{4}$: $q^2=0,-3(4+\sqrt{2}){r_0}^2,3(-4+\sqrt{2}){r_0}^2$\\
		\qquad $\mathrm{AdS}_{5}$: $q^2=(-6+4\sqrt{6}){r_0}^2,-30{r_0}^2,-2(3+2\sqrt{6}){r_0}^2$\\
		\quad \textbf{$2$-form}\\
		\qquad $\mathrm{AdS}_{4}$: $q^2=3(-4+\sqrt{11}){r_0}^2,-3(4+\sqrt{11}){r_0}^2,-9{r_0}^2$\\
		\qquad $\mathrm{AdS}_{5}$: $q^2=2(-9+2\sqrt{14}){r_0}^2,-2(9+2\sqrt{14}){r_0}^2,-18{r_0}^2$\\
		\qquad $\mathrm{AdS}_{6}$: $q^2=0,-50{r_0}^2,-30{r_0}^2$\\
		\quad \textbf{$3$-form}\\
		\qquad $\mathrm{AdS}_{5}$: $q^2=-2(15+2\sqrt{6}){r_0}^2,(-30+4\sqrt{6}){r_0}^2,-6{r_0}^2$\\
		\qquad $\mathrm{AdS}_{6}$: $q^2=\frac{5}{3}(-19+\frac{2^{2/3} 32 }{(13+9i\sqrt{807})^{1/3}}+(26+18i\sqrt{807})^{1/3})r_0^2,$\\ 
		\qquad \qquad \quad$\frac{5}{6}(-38+\frac{2^{2/3} 32(-1\mp i\sqrt{3}) }{(13+9i\sqrt{807})^{1/3}}+(-1\pm i\sqrt{3})(26+18i\sqrt{807})^{1/3})r_0^2$\\
		\qquad $\mathrm{AdS}_{7}$: $q^2=-75{r_0}^2,3(-9+4\sqrt{3}){r_0}^2,-3(9+4\sqrt{3}){r_0}^2$\\
		\hline
	\end{tabular}
\end{table}
\begin{table}[hbt!]
	\caption{The pole-skipping points for the transverse mode of massless $p$-form fields.}
	\label{table_ml_trans}
	\centering
	\begin{tabular}{l}
		\hline
		\textbf{$1$st order} \\
		\;$\mathrm{AdS}_{d+2}$: $\omega=-\frac{1}{2}i(1+d)r_0,\; q^2=-\frac{1}{2}(1+d)(d-2p){r_0}^2$ \\
		\hline
		\textbf{$2$nd order} \\
		\;$\mathrm{AdS}_{d+2}$: $\omega=-i(1+d)r_0,\; q^2=-\left(1+d\right)\left(d-p\pm \sqrt{d-1+\left(p-1\right)^2}\right)r_0^2$ \\
		\hline
		\textbf{$3$rd order} \\
		\;$\omega=-\frac{3}{2}i(1+d)r_0$\\
		\quad \textbf{$1$-form}\\
		\qquad $\mathrm{AdS}_{4}$: $q^2=0,-3(4\pm\sqrt{2}){r_0}^2$\\
		\qquad $\mathrm{AdS}_{5}$: $q^2=-2(9\pm 2\sqrt{14}){r_0}^2, -18{r_0}^2$\\
		\quad \textbf{$2$-form}\\
		\qquad $\mathrm{AdS}_{5}$: $q^2=-2(3\pm 2\sqrt{6}){r_0}^2,-30{r_0}^2$\\
		\qquad $\mathrm{AdS}_{6}$: $q^2=0,-50{r_0}^2,-30{r_0}^2$\\
		\quad \textbf{$3$-form}\\
		\qquad $\mathrm{AdS}_{6}$: $q^2=\frac{5}{3}(-13+\frac{2^{2/3} 56 }{(247+9i\sqrt{3583})^{1/3}}+(494+18i\sqrt{3583})^{1/3})r_0^2,$\\ 
		\qquad \qquad \quad$\frac{5}{6}(-26+\frac{2^{2/3} 56(-1\mp i\sqrt{3}) }{(247+9i\sqrt{3583})^{1/3}}+(-1\pm i\sqrt{3})(494+18i\sqrt{3583})^{1/3})r_0^2$\\
		\qquad $\mathrm{AdS}_{7}$: $q^2=-45{r_0}^2,3(-11+4\sqrt{11}){r_0}^2,-3(11+4\sqrt{11}){r_0}^2$\\
		\hline
	\end{tabular}
\vspace*{3ex}
\end{table}
\begin{table}[hbt!]
	\caption{The pole-skipping points for the longitudinal mode of massive $p$-form fields.}
	\label{table_ms_long}
	\centering
	\begin{tabular}{l}
		\hline
		\textbf{$1$st order} \\
		\;$\mathrm{AdS}_{d+2}$: $\omega=0,\; q^2=-m^2 h(r_0)$ \\
		\hline
		\textbf{$2$nd order} \\
		\;$\mathrm{AdS}_{d+2}$: $\omega=-\frac{1}{2}i(1+d)r_0$,\\
		\qquad \qquad $q^2=-(1/2) (1 + d + 2 m^2 \pm \sqrt{1+d}\sqrt{8 m^2 + (1 + d) (1 + d - 2 p)^2}) r_0^2$ \\
		\hline
		\textbf{$3$rd order} \\
		\;$\mathrm{AdS}_{d+2}$: $\omega=-i(1+d)r_0$, \\
		\qquad \qquad $q^2$ is the root of the following quartic equation of $x$:\\
		\qquad \qquad$-x^4 - 2 x^3 ((1 + d)^2 + 2 m^2) r_0^2 - 
		2 x^2 (-1 - d + d^2 + d^3 - m^2 + 2 d m^2 +$\\ 
		\qquad \qquad$3 d^2 m^2 + 
		3 m^4 + 2 (1 + d)^3 p - 2 (1 + d)^2 p^2) r_0^4 + 
		2 x (-d - 3 d^2 - 2 d^3 +$\\
		\qquad \qquad	
		$ 2 d^4 + 3 d^5 + d^6 - 4 m^2 - 
		4 d m^2 + 4 d^2 m^2 + 4 d^3 m^2 + 5 m^4 + 2 d m^4 - $\\ 
		\qquad \qquad$3 d^2 m^4 -2 m^6 - 4 (1 + d)^3 (-1 + d^2 + m^2) p + 
		4 (1 + d)^2 (-1 + d^2 + $\\
		\qquad \qquad$m^2) p^2) r_0^6 + (-2 - 3 d + d^2 + 3 d^3 +
		d^4 + 4 m^2 + 4 d m^2 - m^4 - 2 (1 + $\\
		\qquad \qquad$d) (-1 + d^2 + m^2) p) (-d -
		d^2 + d^3 + d^4 - 2 m^2 + 2 d^2 m^2 + m^4 - 2 (1 + $\\
		\qquad \qquad$d) (-1 + d^2 + m^2) p) r_0^8=0 $\\
		\hline
	\end{tabular}
\end{table}

\begin{table}[t]
	\caption{The pole-skipping points for the transverse mode of massive $p$-form fields.}
	\label{table_ms_trans}
	\centering
	\begin{tabular}{l}
		\hline
		\textbf{$1$st order} \\
		\;$\mathrm{AdS}_{d+2}$: $\omega=-\frac{1}{2}i(1+d)r_0,\; q^2=-\frac{1}{2}((1+d)(d-2p)+2m^2){r_0}^2$ \\
		\hline
		\textbf{$2$nd order} \\
		\;$\mathrm{AdS}_{d+2}$: $\omega=-i(1+d)r_0,$\\
		\qquad \qquad$q^2=-\left[m^2+\left(d+1\right)\left(d-p \pm \sqrt{d-1+(p-1)^2+\frac{2m^2}{d+1}}\right)\right]{r_0}^2$ \\
		\hline
		\textbf{$3$rd order} \\
		\;$\mathrm{AdS}_{d+2}$: $\omega=-\frac{3}{2}i(1+d)r_0$\\
		\qquad \qquad $q^2$ is the root of the following cubic equation of $x$:\\
		\qquad \qquad$-8 x^3 - 
		4 x^2 (11 d + 11 d^2 + 6 m^2 - 6 (1 + d) p) r_0^2 + 
		2 x (20 d + d^2 - 58 d^3 - $\\ 
		\qquad \qquad$39 d^4 + 32 m^2 - 12 d m^2 - 
		44 d^2 m^2 - 12 m^4 + 4 (1 + d) (-8 + d (3 + 11 d) + $\\ 
		\qquad \qquad$6 m^2) p + 
		4 (1 + d)^2 p^2) r_0^4 + (-48 d - 60 d^2 + 63 d^3 + 69 d^4 - 
		51 d^5 - 45 d^6 - $\\ 
		\qquad \qquad$96 m^2 - 24 d m^2 + 162 d^2 m^2 + 12 d^3 m^2 - 
		78 d^4 m^2 + 64 m^4 + 20 d m^4 - 44 d^2 m^4 - $\\ 
		\qquad \qquad$8 m^6 + 
		2 (1 + d) (3 (1 + d)^2 (16 + d (-28 + 13 d)) + 
		4 (1 + d) (-16 + 11 d) m^2 + $\\ 
		\qquad \qquad$ 12 m^4) p +
		4 (1 + d)^2 (9 d (1 + d) + 2 m^2) p^2 - 24 (1 + d)^3 p^3) r_0^6=0$\\
		\hline
	\end{tabular}
\end{table}

%


\bibliography{reference}
\bibliographystyle{JHEP}

\end{document}